\documentclass{aa}
\usepackage{graphicx}
\usepackage{txfonts}
\usepackage{gensymb,amssymb}
\usepackage{xcolor}
\usepackage[version=3]{mhchem}
\usepackage{caption,subcaption}
\usepackage{multirow}
\usepackage[normalem]{ulem}

\usepackage[colorlinks = true,
            linkcolor = blue,
            urlcolor  = blue,
            citecolor = blue,
            anchorcolor = blue]{hyperref}
\begin{document}

\title{Characterising the molecular line emission in the asymmetric Oph-IRS~48 dust trap: Temperatures, timescales, and sub-thermal excitation}
\titlerunning{Temperatures, timescales, and sub-thermal excitation in the Oph-IRS~48 disk}
\author{Milou Temmink\orcid{0000-0002-7935-7445}\inst{1} \and
        Alice S. Booth\orcid{0000-0003-2014-2121}\inst{2}\thanks{Clay Postdoctoral Fellow} \and
        Margot Leemker\orcid{0000-0003-3674-7512}\inst{3} \and
        Nienke van der Marel\orcid{0000-0003-2458-9756}\inst{1} \and
        Ewine F. van Dishoeck\orcid{0000-0001-7591-1907}\inst{1,4} \and
        Lucy Evans\orcid{0009-0006-1929-3896}\inst{5} \and
        Luke Keyte\orcid{0000-0001-5849-577X}\inst{6} \and 
        Charles J. Law\orcid{0000-0003-1413-1776}\inst{7}\thanks{NASA Hubble Fellowship Program Sagan Fellow} \and
        Shota Notsu\orcid{0000-0003-2493-912X}\inst{8,9} \and
        Karin \"Oberg\orcid{0000-0001-8798-1347}\inst{2} \and
        Catherine Walsh\orcid{0000-0001-6078-786X}\inst{5}}
\institute{Leiden Observatory, Leiden University, PO Box 9513, 2300 RA Leiden, the Netherlands \\
          \email{temmink@strw.leidenuniv.nl} \and
          Center for Astrophysics - Harvard \& Smithsonian, 60 Garden St., Cambridge, MA 02138, USA \and
          Dipartimento di Fisica, Universit\`a degli Studi di Milano, Via Celoria 16, 20133 Milano, Italy \and
          Max-Planck-Institut f\"ur Extraterrestrische Physik, Giessenbachstraße 1, D-85748 Garching, Germany \and
          School of Physics and Astronomy, University of Leeds, LS2 9JT, United Kingdom \and
          Astronomy Unit, School of Physics and Astronomy, Queen Mary University of London, London E1 4NS, UK \and
          Department of Astronomy, University of Virginia, Charlottesville, VA 22904, USA \and
          Department of Earth and Planetary Science, Graduate School of Science, The University of Tokyo, 7-3-1 Hongo, Bunkyo-ku, Tokyo 113-0033, Japan \and 
          Star and Planet Formation Laboratory, RIKEN Cluster for Pioneering Research, 2-1 Hirosawa, Wako, Saitama 351-0198, Japan}
\date{Received 8 September, 2024; accepted 18 November, 2024}
\abstract
{The ongoing physical and chemical processes in planet-forming disks set the stage for planet (and comet) formation. The asymmetric disk around the young star Oph-IRS~48 has one of the most well-characterised chemical inventories, showing molecular emission from a wide variety of species at the dust trap: from simple molecules, such as \ce{CO}, \ce{SO}, \ce{SO_2}, and \ce{H_2CO}, to large complex organics, such as \ce{CH_3OH}, \ce{CH_3OCHO}, and \ce{CH_3OCH_3}. One of the explanations for the asymmetric structure in the disk is dust trapping by a perturbation-induced vortex.}
{We aim to constrain the excitation properties of the molecular species \ce{SO_2}, \ce{CH_3OH}, and \ce{H_2CO}, for which we have used, respectively, 13, 22, and 7 transitions of each species. We further characterise the extent of the molecular emission, which differs among molecules, through the determination of important physical and chemical timescales at the location of the dust trap. We also investigate whether the anticyclonic motion of the potential vortex influences the observable temperature structure of the gas.}
{Through a pixel-by-pixel rotational diagram analysis, we create maps of the rotational temperatures and column densities of \ce{SO_2} and \ce{CH_3OH}. To determine the temperature structure of \ce{H_2CO}, we have used line ratios of the various transitions in combination with non-LTE RADEX calculations. The timescales for freeze-out, desorption, photodissociation, and turbulent mixing at the location of the dust trap are determined using an existing thermochemical model.}
{Our rotational diagram analysis yields temperatures of $T$=54.8$\pm$1.4 K (\ce{SO_2}) and $T$=125.5$^{+3.7}_{-3.5}$ K (\ce{CH_3OH}) at the emission peak positions of the respective lines. As the \ce{SO_2} rotational diagram is well characterised and points towards thermalised emission, the emission must originate from a layer close to the midplane where the gas densities are high enough. The rotational diagram of \ce{CH_3OH} is, in contrast, dominated by scatter and subsequent non-LTE RADEX calculations suggest that both \ce{CH_3OH} and \ce{H_2CO} must be sub-thermally excited higher up in the disk ($z/r\sim$0.17-0.25). For \ce{H_2CO}, the derived line ratios suggest temperatures in the range of $T\sim$150-350 K. The \ce{SO_2} temperature map hints at a potential radial temperature gradient, whereas that of \ce{CH_3OH} is nearly uniform and that of \ce{H_2CO} peaks in the central regions. We, however, do not find any hints of the vortex influencing the temperature structure across the dust trap. The longer turbulent mixing timescale, compared to that of photodissociation, does provide an explanation for the expected vertical emitting heights of the observed molecules. On the other hand, the short photodissociation timescales are able to explain the wider azimuthal molecular extent of \ce{SO_2} compared to \ce{CH_3OH}. The short timescales are, however, not able to explain the wider azimtuhal extent of the \ce{H_2CO} emission. Instead, it can be explained by a secondary reservoir that is produced through the gas-phase formation routes, which are sustained by the photodissociation products of, for example, \ce{CH_3OH} and \ce{H_2O}.}
{Based on our derived temperatures, we expect \ce{SO_2} to originate from deep inside the disk, whereas \ce{CO} comes from a higher layer and both \ce{CH_3OH} and \ce{H_2CO} emit from the highest emitting layer. The sub-thermal excitation of \ce{CH_3OH} and \ce{H_2CO} suggests that our derived (rotational) temperatures underestimate the kinetic temperature. Given the non-thermal excitation of important species, such as \ce{H_2CO} and \ce{CH_3OH}, it is important to use non-LTE approaches when characterising low-mass disks, such as that of IRS~48. Furthermore, for the \ce{H_2CO} emission to be optically thick, as expected from an earlier derived isotopic ratio, we suggest that the emission must originate from a small radial ``sliver'' with a width of $\sim$10 au, located at the inner edge of the dust trap. }


\maketitle

\section{Introduction}
Planets form in the disks surrounding newly formed stars. Their atmospheric composition, as well as the composition of comets, is determined by the chemical makeup of these planet-forming disks, which itself is the result of the combination of ongoing chemical processes and the inheritance of molecular species from the earlier star-formation stages (e.g. \citealt{OB21, ObergEA23}). The Atacama Large Millimeter/submillimeter Array (ALMA) allows us to characterise the chemical inventory of planet-forming disks with unprecedented sensitivity. The sensitivity enables us to detect multiple transitions of the same species that span a range of excitation conditions, even those with weaker line strengths. Detecting multiple transitions allows one, under the assumption of local thermal equilibrium (LTE), to directly infer both the gas thermal and excitation properties through a rotational diagram analysis (e.g. \citealt{GL99}). The use of multiple lines of the same molecule in order to constrain the physical structure of disks has a long history (see, e.g., \citealt{vZadelhoffEA01,DartoisEA03,LeemkerEA22}). Such studies have also highlighted the potential importance of non-LTE excitation in the upper layers of the disks (e.g., \citealt{vZadelhoffEA03,ThiEA04,WoitkeEA09}). \citet{LoomisEA18} presented a rotational diagram analysis for a protoplanetary disk with the detection of \ce{CH_3CN} in the TW~Hya disk, and rotational diagrams have also been used for a variety of molecules in other disks (e.g. \citealt{BergnerEA18,GuzmanEA18,PeguesEA20,LeGalEA19,LeGalEA21,CleevesEA21,MunozRomeroEA23,TemminkEA23,RampinelliEA24}). Using the derived excitation properties, unresolved observations can be linked to the radial extent and vertical height of the molecular emission in the disk and, therefore, the chemical origin of the species may be inferred (e.g, \citealt{CarneyEA17,SalinasEA17,CarneyEA18,IleeEA21,TvSEA21,MunozRomeroEA23,HernandezVeraEA24}). \\
\indent One of the disks with the best-studied chemical inventory is that of Oph-IRS~48 (IRS~48 hereafter), a Herbig A0 star ($M\sim$2.0 M$_\odot$ and $L\sim$17.8 L$_\odot$; \citealt{BrownEA12,FvdM20}) located in the Ophiuchus star-forming region (at a distance of 135 pc; \citealt{GC18}) with many simple and complex organic molecules (COMs) detected \citep{vdMarelEA14,vdMarelEA21,BoothEA21,BrunkenEA22,LeemkerEA23,BoothEA24}, with the latter defined as molecules with $\geq$6 atoms of which at least one is carbon \citep{HvD09}. IRS~48 is well known for its asymmetric dust trap, in which the large, millimetre-sized dust grains have been captured mostly on the southern side of the disk \citep{vdMarelEA13,YangEA23}. The molecular emission originates from approximately the same location as the asymmetric dust trap, suggesting that the sublimation of the icy mantles of pebbles plays an important role in setting the observable gas content. The dust asymmetry is potentially the result of a vortex, introduced due to an instability in the disk. The Rossby Wave Instability (RWI), for which the theoretical groundwork in disks was first described by \citet{LovelaceEA99} and \citet{LiEA00}, is often thought to be the cause of the vortex. Various studies (see, for example, \citealt{LyraEA09,CrnkovicEA15,OwenEA17,RaettigEA21,RegalyEA21}) have shown that these vortices can trap significant amounts of dust if it has an anticyclonic rotation, can have large dust-to-gas ratios and, therefore, can act as the formation sites of planetary and/or cometary bodies. \\ 
\indent One curious matter of these asymmetric dust traps is the temperature structure of the gas. Normally, the temperature of disks decreases radially outwards, i.e. the further away from the host star the lower the temperature, and vertically downwards, i.e. lower temperatures are reached at the midplane of the disk, unless there is significant heating due to viscous accretion in the midplane (e.g. \citealt{DAlessioEA98}, \citealt{WoitkeEA09}, and \citealt{HarsonoEA15}). Intuitively, the vortex may add additional complexity to the temperature structure, as its motions may transfer hot gas out of view of the host star, towards larger radii, and cold gas into view. Additionally, direct UV irradiation can heat the gas at cavity edges \citep{BrudererEA13,BrudererEA14,LeemkerEA23}. Consequently, the temperature may not decrease radially in these kinds of disks. In this paper, we study the temperature structure of the gas, as well as the origins and distribution of different chemical species, in the IRS~48 disk using the plethora of observed \ce{SO_2}, \ce{CH_3OH}, and \ce{H_2CO} transitions, which have first been reported by \citet{vdMarelEA21}, \citet{BoothEA21}, and \citet{BoothEA24}. We use complementary observations of the optically thick \ce{^{13}CO} isotopologue \citep{vdMarelEA16,LeemkerEA23} to further characterise the temperature and the vertical distribution of the molecular species. \\
\indent This paper is structured as follows: in Section \ref{sec:TargetData} we describe our target and the used data, while our analysis is explained in Section \ref{sec:Analysis}. Our results can be found in Section \ref{sec:Results}, which are further discussed in Section \ref{sec:Discussion}. Section \ref{sec:Timescales} contains an analysis of important physical and chemical timescales that aid in the explanation of the observed molecular signatures. Finally, our conclusions are summarised in Section \ref{sec:Conclusions}.

\section{Data} \label{sec:TargetData}
\begin{figure*}[h!]
    \centering
    \includegraphics[width=\textwidth]{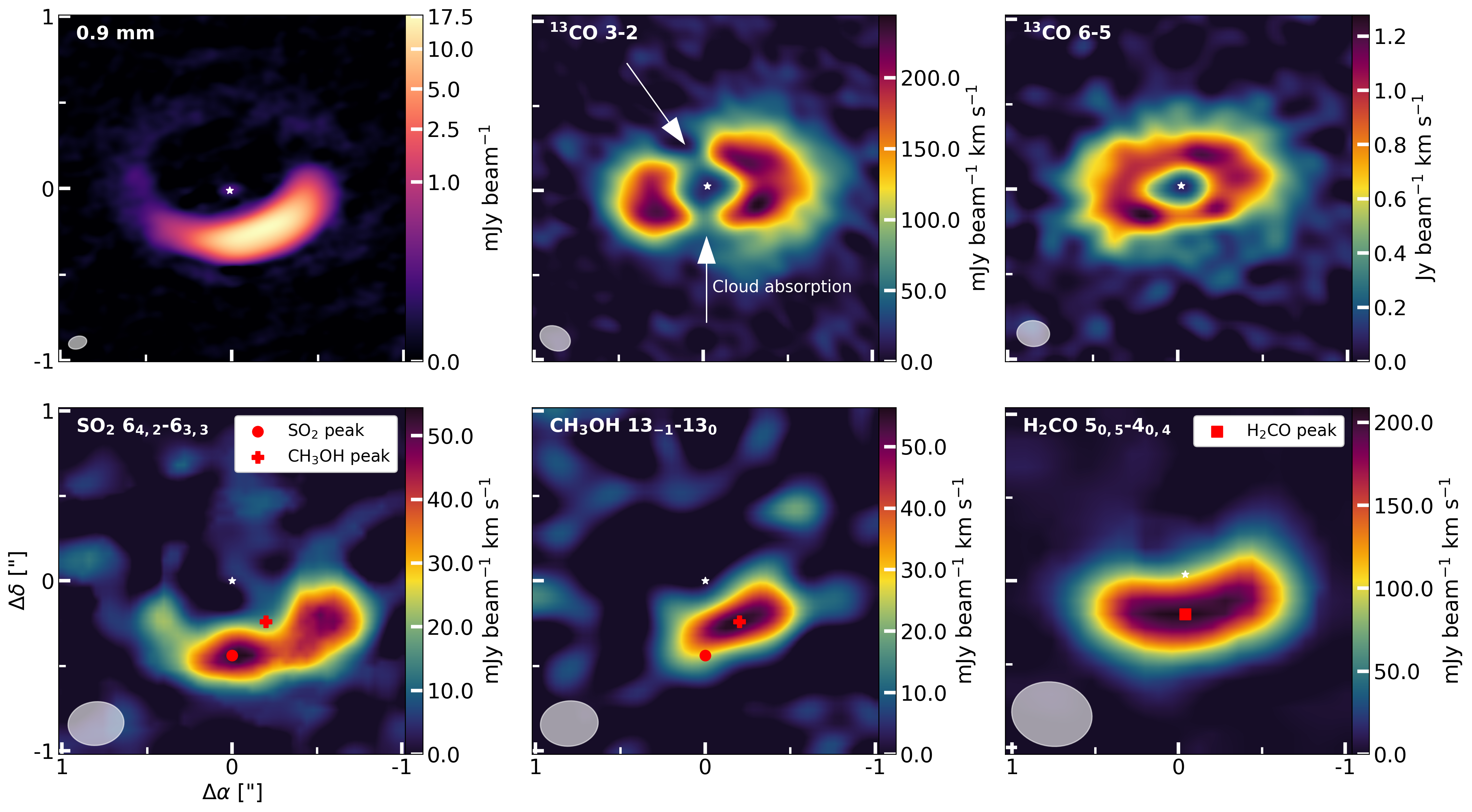}
    \caption{Dust continuum (top left; \citealt{YangEA23}) and moment-0 maps of the \ce{^{13}CO} $J$=3-2 (top centre; \citealt{LeemkerEA23}) and $J$=6-5 (top right; \citealt{vdMarelEA16}), and the \ce{SO_2} $J$=6$_{4,2}$-6$_{3,3}$ (bottom left), \ce{CH_3OH} $J$=13$_{-1}$-13$_0$ (bottom centre), and \ce{H_2CO} $J$=5$_{0,5}$-4$_{0,4}$ (bottom right) transitions. The white star in the centre indicates the approximate location of the host star, whereas the resolving beam is indicated in the lower left. The red circle and cross provide, respectively, the peak flux position of the \ce{SO_2} and \ce{CH_3OH} transitions in the bottom left and centre plots, whereas the red square denotes the peak flux position of the \ce{H_2CO} transition in the bottom right plot. In addition, the arrows in the \ce{^{13}CO} $J$=3-2 transition indicate where the emission is affected by the cloud absorption.}
    \label{fig:M0s}
\end{figure*}

\indent Throughout this work, we use Band 7 ALMA data of IRS~48 observed in Cycle 8 (2021.1.00738.S, PI: A. S. Booth). We refer to \citet{BoothEA24} for a description of the data reduction, self-calibration, and continuum subtraction. The observations consist of 8 spectral windows, covering transitions at frequencies between $\sim$336.9 and $\sim$356.9 GHz at a spectral resolution of $\sim$0.84 km s$^{-1}$. We detect 13 and 22 clean (i.e. not blended with other molecules) transitions of \ce{SO_2} and \ce{CH_3OH}, respectively. Through a pixel-by-pixel rotational diagram analysis (pixel sizes of 0.04", beam sizes of 0.34"$\times$0.26"), we investigate the temperature of the IRS~48 disk, as probed by these molecules. Throughout our analysis, we used all pixels within a radius of 1.5" of the host star as seen within the disk coordinates (i.e., accounting for the disk's inclination, $i$=50\degree, and position angle, $PA$=100\degree). In addition, we use the lower angular resolution ($\sim$0.46"), Cycle 5 Band 7 observations of \ce{H_2CO} presented in \citet{vdMarelEA21}. \\
\indent The integrated flux, or moment-0, maps of the \ce{SO_2} 6$_{4,2}$-6$_{3,3}$, \ce{CH_3OH} 13$_{-1}$-13$_0$, and \ce{H_2CO} 5$_{0,5}$-4$_{0,4}$ transitions are displayed in Figure \ref{fig:M0s}, together with high-resolution, 0.9~mm dust observations \citep{YangEA23} and the \ce{^{13}CO} $J$=3-2 \citep{LeemkerEA23} and $J$=6-5 \citep{vdMarelEA16} transitions. These transitions of \ce{SO_2}, \ce{CH_3OH}, and \ce{H_2CO} are among the brightest ones observed in the disk and, therefore, their peak position provide good insights into the excitation conditions at the positions in the disk where these molecules are strongly detected. Immediately clear from the moment-0 maps are the differences in the azimuthal extent of the molecular species: the \ce{CH_3OH} is very compact at the location of the dust trap, whereas the emission of both \ce{SO_2} and \ce{H_2CO} is spread out over a larger azimuthal extent, up to approximately quarter of an orbit. The \ce{^{13}CO} transitions, in particular the $J$=3-2 one, are affected by cloud absorption. Within the images, we have indicated the pixel position of the peak flux and throughout this paper we show examples for these peak pixel positions.

\section{Methods} \label{sec:Analysis}
\subsection{Identifying detected transitions} \label{sec:LI}
The various \ce{SO_2}, \ce{CH_3OH}, and \ce{H_2CO} transitions are, for each separate pixel, identified using the `Leiden Atomic and Molecular Database' (LAMDA; \citealt{CDMS,LAMDA,WF13,RF10b,BalancaEA16}). We limit our search to transitions with Einstein-A coefficients of $A_\textnormal{ul} \geq 10^{-6}$ s$^{-1}$ and upper level energies of $E_\textnormal{up}\leq$500 K, as lines with lower Einstein-A coefficients and higher upper level energies are not detected. For each pixel and spectral window, we obtain the corresponding spectrum in units of Jy beam$^{-1}$. The spectra extracted for the pixel where the moment-0 maps peak (highlighted in Figure \ref{fig:M0s}) are shown in Figures \ref{fig:SO2-Lines}, \ref{fig:CH3OH-Lines}, and \ref{fig:H2CO-Lines} for \ce{SO_2}, \ce{CH_3OH}, and \ce{H_2CO}, respectively. Within each spectrum, we consider all transitions with peak fluxes above three times the root-mean-square (RMS) as detections. As we retrieve spectra for each pixel and all spectral windows, we determine the RMS in each spectrum using an automated approach consisting of various steps: first, the residual variation in the continuum-subtracted baseline is determined using an iterative, 2$\sigma$-outlier masking process and a Savitzky-Golay filter \citep{SavGol}. The Savitzky-Golay filter allows one to smooth the data using a low-order polynomial over a subset of the data, without altering the underlying shape of the spectrum. We filtered the data using third-order polynomials that smoothed the data every 100 data points. Secondly, the standard deviation (SD) of the residuals, after subtracting the final Savitzky-Golay filter from the 2$\sigma$-outlier masked spectrum, is obtained. Finally, we identify and mask all emission lines from the full spectrum, by removing all emission lines that have fluxes above 3$\times$ the aforementioned SD. The RMS is, subsequently, obtained from the resulting line-free spectrum. Overall, we find RMS values of $\sim$1 mJy~beam$^{-1}$ for the Cycle 8 observations, very similar to the values listed in Table 6 of \citet{BoothEA24}. The RMS varies between $\sim$1.0-1.5 mJy~beam$^{-1}$ for the Cycle 5 observations, in agreement with the value listed in \citet{vdMarelEA21} of $\sim$1.2 mJy~beam$^{-1}$. \\
\indent From our list of detected transitions, we have removed transitions which are blended with other transitions of the same molecule with different excitation conditions and/or blended with transitions of other molecules. However, we have kept the transitions that are blended with the transitions of the same molecule that have the same Einstein-A coefficient and upper level energy. This allowed us to include two pairs of \ce{CH_3OH} transitions. As both transitions have the same excitation conditions and, therefore, are expected to contribute equally to the observed line, we assume that half of the observed line flux can be attributed to each of the transitions. Furthermore, through trial and error and visual inspection of the image cubes and line profiles, we have removed additional \ce{SO_2} transitions for which we could not confidently distinguish between emission and 3$\sigma$ noise spikes, i.e. features that appear only locally and do not follow the emission morphologies (i.e. do not follow the same pattern, such as Keplerian rotation, or display different line profiles) seen for stronger transitions. All detected transitions are listed in Table \ref{tab:PeakDetections}, whereas the blended transitions or those dominated by noise are listed in Table \ref{tab:RemTrans}. 

\subsection{Integrated intensity and rotational diagram analysis} \label{sec:Intensity}
We infer the excitation conditions (rotational temperature and column density) using a rotational diagram analysis. We obtain integrated fluxes by fitting Gaussian line profiles. As the flux from individual pixels is not independent, we consider the beam area to be the emitting area in our calculations. In addition, we follow the method described by, for example, \citet{BanzattiEA12} to create model intensities, which, subsequently, are fitted to inferred values. In the calculation of the optical depth, we have used a line width of $\Delta V_\textnormal{line}$=$\sqrt{\Delta V_\textnormal{thermal}^2+\Delta V_\textnormal{turbulence}^2}$=0.2 km~s$^{-1}$. This line width is a combination of the thermal line width (expected to be on the order of $\sim$0.15 km~s$^{-1}$) and the intrinsic width due to local turbulence, which can also reach values of $\sim$0.1 km~s$^{-1}$ (see, e.g., \citealt{PanequeEA24}). Additionally, we have calculated 3$\sigma$ upper limits, following the description of \citet{CarneyEA19}, for the pixels for which we have not created a rotational diagram.

\section{Results} \label{sec:Results}
\begin{figure*}[ht!]
    \centering
    \includegraphics[width=\textwidth]{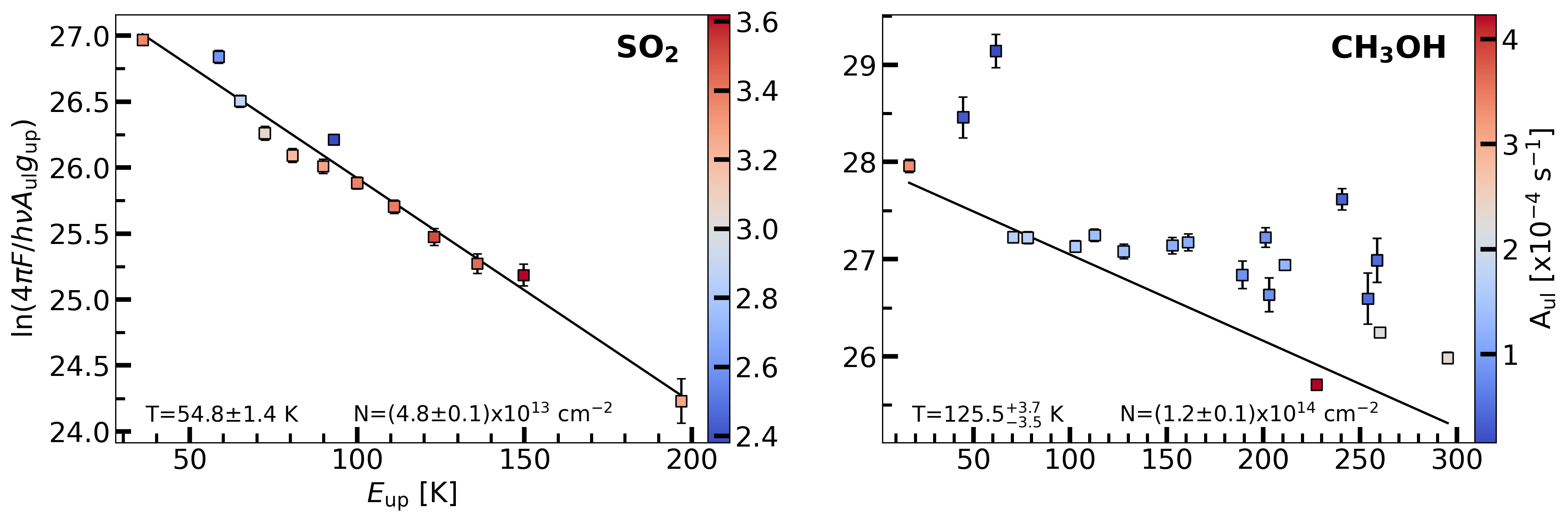}
    \caption{Rotational diagrams for \ce{SO_2} (left) and \ce{CH_3OH} (right) at the peak flux positions of their 6$_{4,2}$-6$_{3,3}$ and 13$_{-1}$-13$_0$ transitions, respectively. The colour bar indicates the value for $A_\textnormal{ul}$ ($\times$10$^{-4}$) of the transitions. }
    \label{fig:RDAs}
\end{figure*}
\begin{figure*}[ht!]
    \centering
    \includegraphics[width=\textwidth]{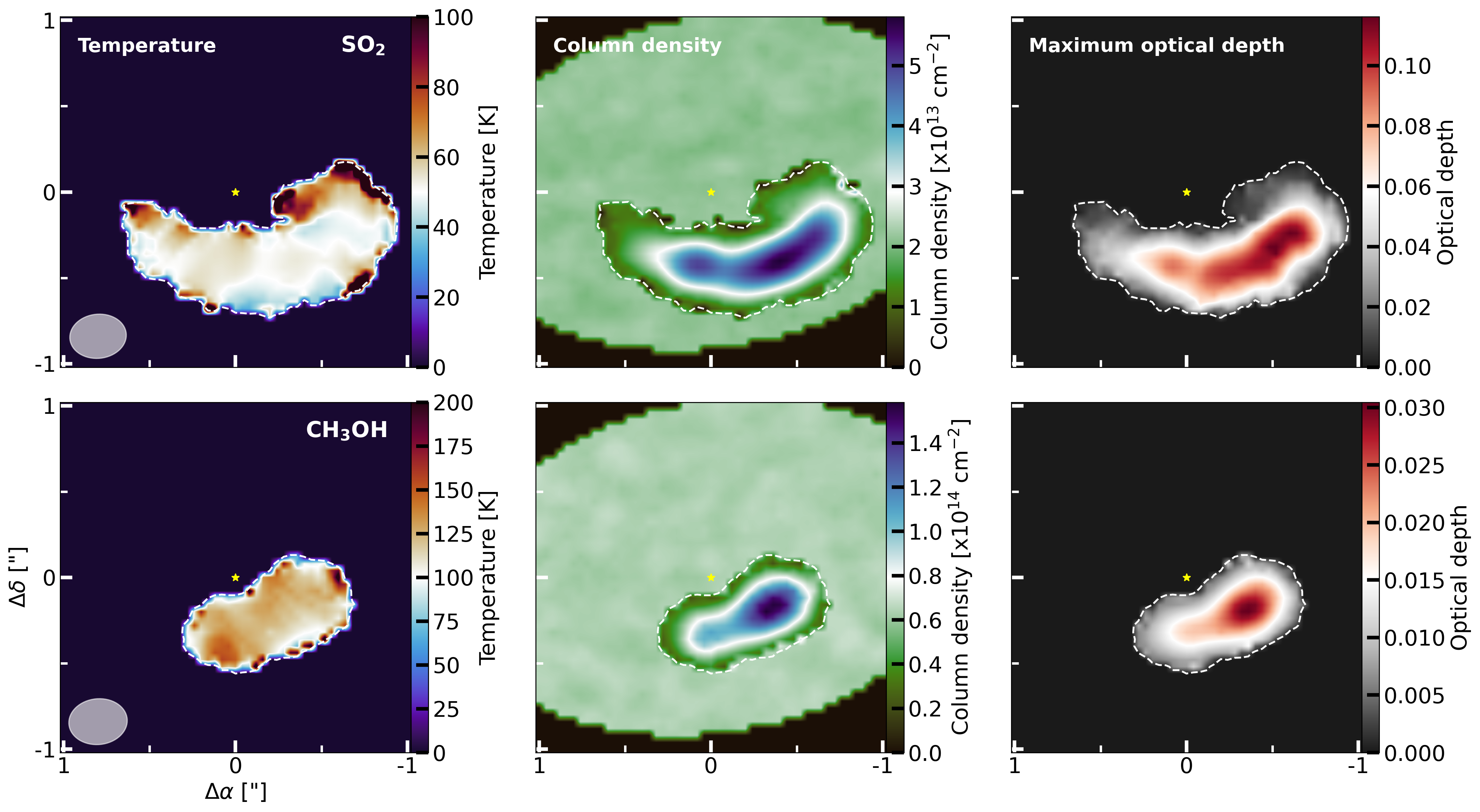}
    \caption{Temperature (left column), column density (middle column), and maximum optical depth (right column) maps for \ce{SO_2} (top) and \ce{CH_3OH} (bottom).}
    \label{fig:Maps}
\end{figure*}

Below we present the rotational diagrams (Figure \ref{fig:RDAs}) for \ce{SO_2} and \ce{CH_3OH}. In addition, we show the resulting temperature, column density, and optical depth maps (Figures \ref{fig:Maps} and \ref{fig:RatioMaps} and line profiles (Figures \ref{fig:SO2-Lines}, \ref{fig:CH3OH-Lines}, and \ref{fig:H2CO-Lines}) retrieved for the molecules at the peak position of their $J$=6$_{4,2}$-6$_{3,3}$, $J$=13$_{-1}$-13$_0$, and $J$=5$_{0,5}$-4$_{0,4}$ transitions, respectively. These peak positions are highlighted in Figure \ref{fig:M0s}. Our analysis assumes that the molecular emission is optically thin, for a discussion on the potential effects of optically thick emission, see Section \ref{sec:ODES}

\subsection{\ce{SO_2}}
At the peak position of the integrated intensity map of the $J$=6$_{4,2}$-6$_{3,3}$ transitions of \ce{SO_2}, we detect 13 isolated transitions (see also Table \ref{tab:PeakDetections}). The rotational diagram for this position is displayed in the left panel of Figure \ref{fig:RDAs}, while the line profiles are shown in Figure \ref{fig:SO2-Lines}. The left-side of Figure \ref{fig:RDAs-Ratios} displays the intensity ratios between the observations and the modelled intensities. We find a well-constrained rotational diagram and derive a rotational temperature of $T$=54.8$\pm$1.4 K and a peak total column density of $N$=(4.8$\pm$0.1)$\times$10$^{13}$ cm$^{-2}$. The upper limits on the column density, for pixels with an insufficient number of detected transitions, range between $\sim$2.0$\times$10$^{13}$ and $\sim$2.4$\times$10$^{13}$ cm$^{-2}$, which are a factor of $\sim$2 lower than found at the peak position. We also find, for all pixel positions, a maximum optical depth of $\tau\sim$0.1, suggesting that the \ce{SO_2} emission is optically thin. The optical depth is further discussed in Section \ref{sec:ODES}. The final maps (temperature, column density, and optical depth) are displayed in the left panel of Figure \ref{fig:Maps}. Given the low inferred uncertainties on the temperatures (on the order of a few K, see Figure \ref{fig:TUncertainty}), the hotter temperature ($\sim$70 K) at the inner edge is significantly higher than that at the outer edge ($\sim$40 K). The higher temperatures inferred at some of the edge pixels are likely due to fewer transitions detected and weaker line fluxes, and are, therefore, more uncertain (see also Figure \ref{fig:TUncertainty}). \\
\indent The well-behaved rotational diagram of \ce{SO_2} points towards thermalised emission. As the gas densities necessary for the \ce{SO_2} emission to become thermalised are only found in a layer just above the midplane, we suggest that the emission must originate from this layer deep inside the disk (see also Section \ref{sec:VS}), which is also consistent with the derived rotational temperature.

\subsection{\ce{CH_3OH}}
For \ce{CH_3OH}, we observe 22 transitions at the peak position of the integrated intensity map of the $J$=13$_{-1}$-13$_0$ transition that are either unblended or blended with other \ce{CH_3OH} lines with similar parameters (see also Table \ref{tab:PeakDetections}). The rotational diagram and the ratios between the observations and modelled intensities are shown in the right panels of Figures \ref{fig:RDAs} and \ref{fig:RDAs-Ratios}, the line profiles are displayed in Figure \ref{fig:CH3OH-Lines}, and the final maps are shown in the lower row of Figure \ref{fig:Maps}. Using these transitions, we derive a rotational temperature of $T$=125.5$^{+3.7}_{-3.5}$ K and a total column density of $N$=(1.2$\pm$0.1)$\times$10$^{14}$ cm$^{-2}$ at the peak position. The upper limits on the column density are on the order of $\sim$5.8$\times$10$^{13}$ cm$^{-2}$ to $\sim$6.9$\times$10$^{13}$ cm$^{-2}$, a factor of $\sim$2 lower than the peak value. We find a maximum optical depth of $\tau\sim$0.03, suggesting that the \ce{CH_3OH} emission is optically thin. In contrast with that of \ce{SO_2}, the rotational diagram of \ce{CH_3OH} shows a large scatter between the data points, this is further discussed in Section \ref{sec:STE}. 

\subsection{\ce{H_2CO}} \label{sec:Res-H2CO}
\begin{figure*}[h!]
    \centering
    \includegraphics[width=\textwidth]{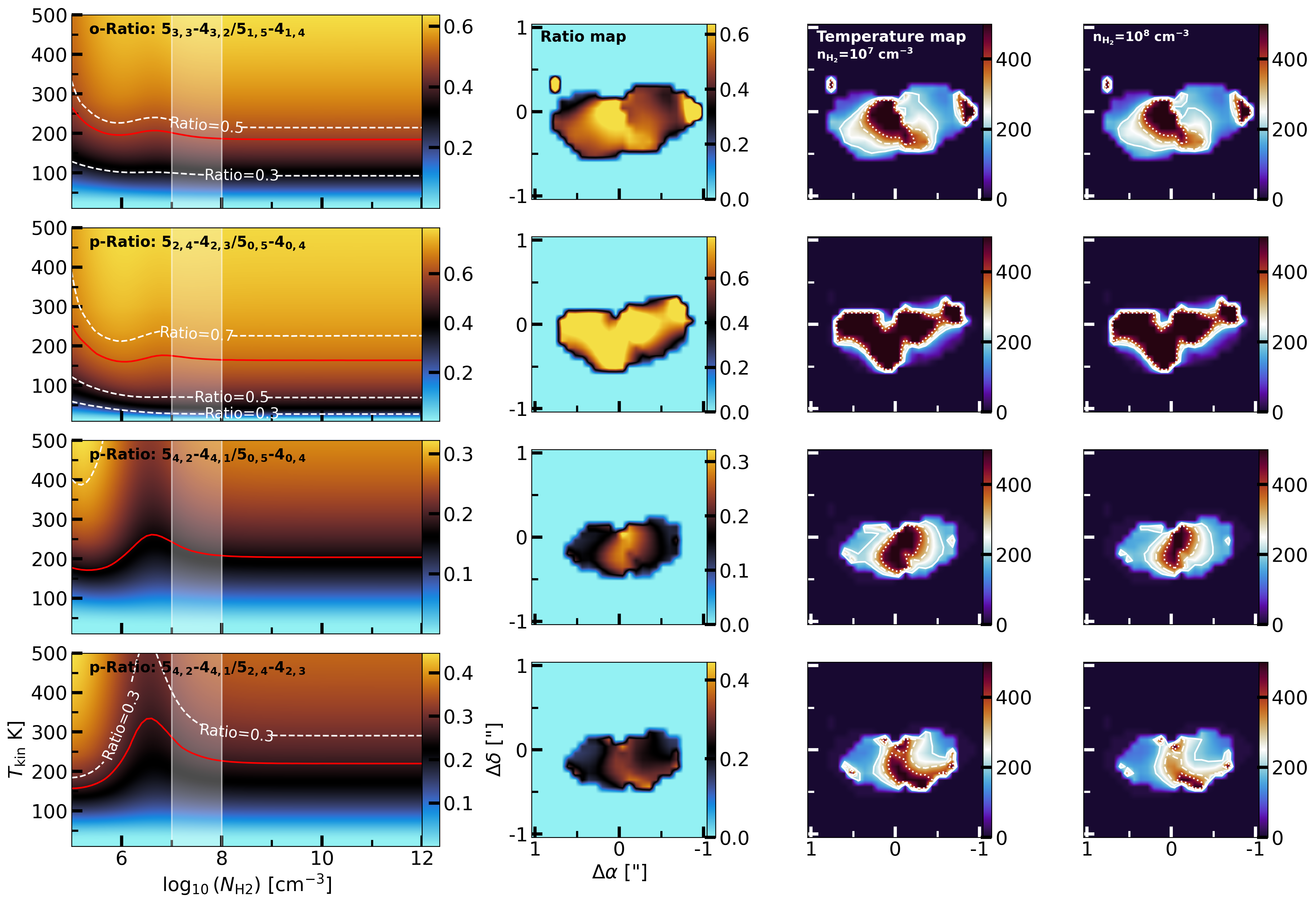}
    \caption{Line ratios for the different pairs of \ce{H_2CO} transitions. The left panels show the RADEX calculations, with the colour map indicating the various ratios. The white contours indicate the values of 0.3, 0.5, and 0.7, whereas the red contour indicates the disk integrated ratio from \citet{vdMarelEA21}. The panels in the second column indicate the ratios derived for the pixels where both transitions are detected, using the colour scheme as for the left panels. The final two columns display the derived temperatures for gas densities of $n_\textnormal{\ce{H_2}}$=10$^{7}$ cm$^{-3}$ (representative of the disk's atmosphere, $z/r\gtrsim$0.2) and 10$^8$ cm$^{-3}$ (representative of the disk's midplane). The white solid, dashed, and dotted contours indicate temperatures of 200 K, 300 K, and 400 K, respectively.}
    \label{fig:RatioMaps}
\end{figure*}
The pixel-by-pixel rotational diagram analysis for the 7 observed \ce{H_2CO} transitions in Cycle 5 \citep{vdMarelEA21} proved to be more difficult and we were not able to retrieve reliable values for the rotational temperature and column density, likely due to a combination of the optical depth (discussed later in this section), a mixture of the ortho- and para-transitions, and sub-thermal excitation (see Section \ref{sec:STE}). In order to obtain a pixel-by-pixel temperature map, we derived the temperatures using various line ratios. Similar to \citet{vdMarelEA21}, we use RADEX to obtain line ratios for various values of the kinetic temperature ($T_\textnormal{kin}$ in K) and the gas density ($n_\textnormal{\ce{H_2}}$ in cm$^{-3}$). Following \citet{vdMarelEA21} we used a column density of $N_\textnormal{\ce{H_2CO}}$=10$^{14}$ cm$^{-2}$ in the calculations, at which the emission should be optically thin. Additionally, we have used the default line width of 1 km~s$^{-1}$ within the RADEX calculations. More details on the calculations can be found in Section \ref{sec:STE}. We determine the temperature, using a simple minimisation of the ratios obtained through RADEX with the inferred ratios from the observations, for gas densities of $n_\textnormal{\ce{H_2}}$=10$^7$ cm$^{-3}$ and 10$^8$ cm$^{-3}$, which, respectively, correspond to the disk's atmosphere and the midplane. \\
\indent We have determined the line ratios for one pair of ortho-transitions (5$_{3,3}$-4$_{3,2}$/5$_{1,5}$-4$_{1,4}$) and three pairs of para-transitions (5$_{2,4}$-4$_{2,3}$/5$_{0,5}$-4$_{0,4}$, 5$_{4,2}$-4$_{4,1}$/5$_{0,5}$-4$_{0,4}$, and 5$_{4,2}$-4$_{4,1}$/5$_{2,4}$-4$_{2,3}$). The second and third pairs of the para-transitions allow us to investigate how the ratios change for transitions with different upper level energies, as the line excitation and, therefore, the observed flux and ratios are temperature sensitive. The ratios and inferred temperature maps are displayed in Figure \ref{fig:RatioMaps}. We highlight our findings for each line pair separately in Section \ref{sec:H2CO-LRs} of the Appendix. \\
\indent Overall, we can infer from the ratios that the temperatures of \ce{H_2CO} across the dust trap can be expected to be on the order of $T\sim$150-350 K, with the highest temperatures found in the centre of the dust trap. As discussed above, this is generally similar to, albeit slightly higher than, the ratios derived from the disk-integrated values of \citet{vdMarelEA21}. Such high temperatures are not unexpected given the high upper level energies of some of the \ce{H_2CO} transitions detected in the disk of IRS~48 (i.e. the $J$=5$_{4,2}$-5$_{4,1}$ and $J$=5$_{4,1}$-5$_{4,0}$ transitions with $E_\textnormal{up}$=240.7 K or the warm transition of $J$=9$_{1,8}$-8$_{1,7}$ with $E_\textnormal{up}$=174 K detected in \citealt{vdMarelEA14}). Compared to the other molecules, \ce{SO_2} and \ce{CH_3OH}, we find that the temperatures are significantly higher than those of \ce{SO_2}, but are similar to and/or higher than those of \ce{CH_3OH}. These high temperatures indicate that the emission must come from high up in the disk, as suggested before by \citet{vdMarelEA21}. Therefore, the temperatures derived assuming a gas density of $n_\textnormal{\ce{H_2}}$=10$^7$ cm$^{-3}$ should provide the best constraints. These are, however, also more limited by the ratios and the involved temperature sensitivity of the transitions. For the remainder of this work, we assume that the temperature derived for \ce{H_2CO} is at least $\sim$150 K. \\
\indent For a temperature of $T$=150 K, a column density of $N$=10$^{15}$ cm$^{-2}$, and a gas density of $n_\textnormal{\ce{H_2}}=$10$^{7}$ cm$^{-3}$, the RADEX calculations yield a maximum optical depth of $\tau\sim$1.9, suggesting that some of the transitions may be (moderately) optically thick in the outer edges of the dust trap (i.e. the regions where probe temperatures of $T_\textnormal{rot}\sim$150 K, see Figure \ref{fig:RatioMaps}). For the same conditions, but a temperature of $T$=350 K, we infer a maximum optical depth of $\tau\sim$0.6, suggesting that the emission in the central regions are more optically thin.

\section{Discussion} \label{sec:Discussion}
In the following sections, we discuss our results. We start by discussing the well-behaved rotational diagram of \ce{SO_2}, which is in contrast with that for \ce{CH_3OH} which is dominated by large scatter. This points towards \ce{SO_2} being thermally excited, while \ce{CH_3OH} is sub-thermally excited (Section \ref{sec:STE}). In Section \ref{sec:TempStruc}, we discuss the temperature and resulting vertical structures of the disk, whereas the optical depth and radial structure are discussed in Section \ref{sec:ODES}. Finally, we highlight potential follow-up observations in Section \ref{sec:FutureObs}, which are needed to fully resolve the molecular emission and to infer more information on the column densities.

\subsection{Sub-thermal excitation} \label{sec:STE}
Given the well behaved rotational diagram of \ce{SO_2}, characterised by very limited scatter, we expect the emission to be (close to) thermalised. In stark contrast with this, is the large scatter observed in the rotational diagram of \ce{CH_3OH}, which is likely due to the effects of sub-thermal excitation. As shown in Figure 6 of \citet{JohnstoneEA03}, higher energy lines of \ce{CH_3OH} have higher critical densities and are, subsequently, not thermalised at relatively low densities (i.e. $n$=10$^7$ cm$^{-3}$). A rotational diagram can still yield a single excitation temperature using a rotational diagram, but this value is much lower than the kinetic one, because the levels are not thermalised. By increasing the density (to, for example, $n$=10$^9$ cm$^{-3}$) in their RADEX calculations, \citet{JohnstoneEA03} found that the lines became thermalised tended towards the physical input conditions. In particular, the derived rotational temperature started to represent the input kinetic temperature. \\
\begin{figure}[ht!]
    \centering
    \includegraphics[width=\columnwidth]{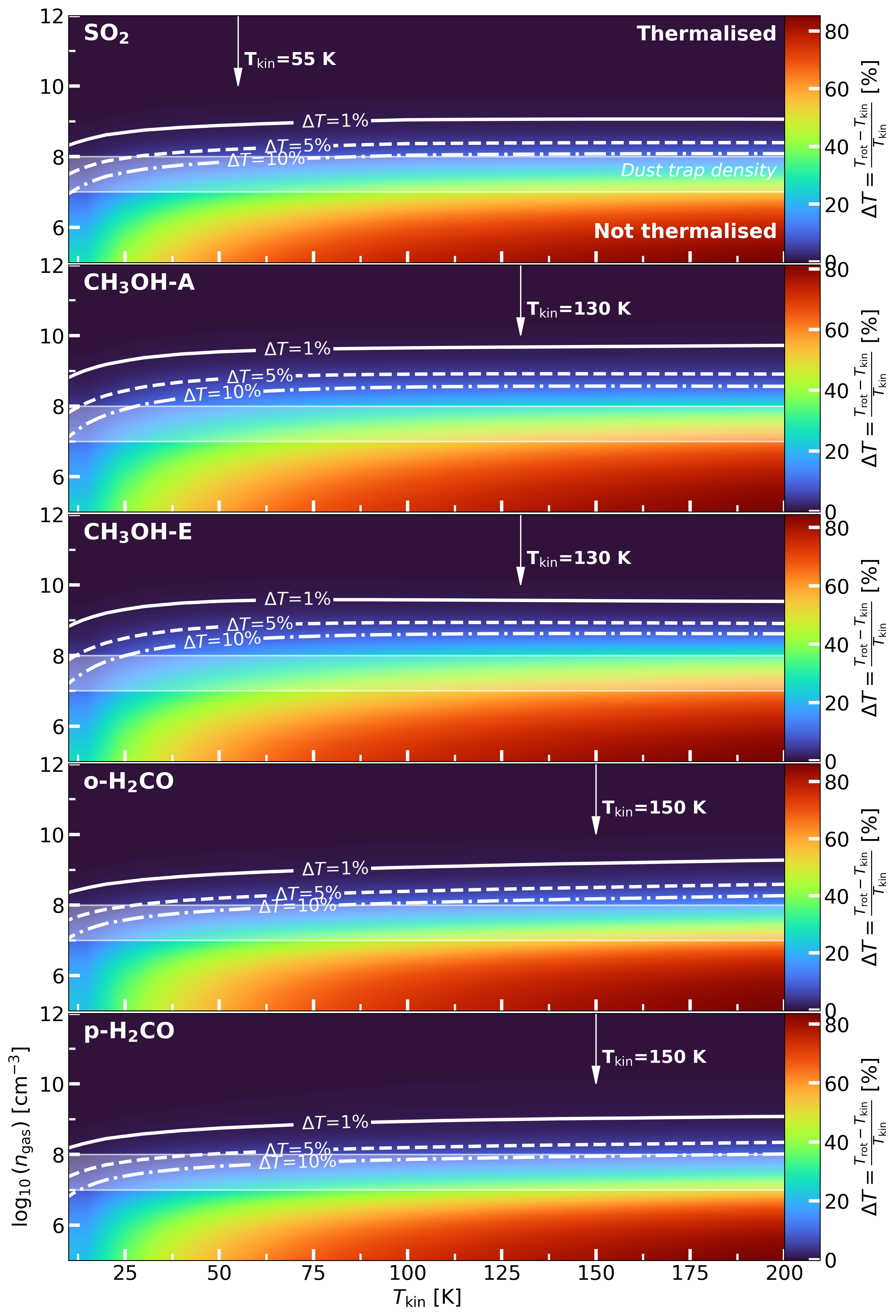}
    \caption{RADEX grids for \ce{SO_2} (top panel), A- and E-type \ce{CH_3OH} (second and third panels), and ortho- and para-\ce{H2CO} (bottom two panels) showing the percentile difference between the kinetic and rotational temperature ($\Delta T$), as a measure of thermalised transitions. The contours (solid, dashed, and dashed-dotted) in each panel indicates when $\Delta T$=1\%, 5\%, and 10\%, respectively. Above these contours the lines will be thermalised. The white shaded area indicates the gas density of the dust trap according to the DALI model \citep{LeemkerEA23}. Finally, the arrows indicate the derived excitation temperatures of $T\sim$55 K for \ce{SO_2} and $T\sim$130 K for \ce{CH_3OH}. For \ce{H_2CO} we show the rotational temperature of 150 K, the lower value of the inferred temperature range.}
    \label{fig:RADEX}
\end{figure}
\indent We have used RADEX \citep{RADEX} to investigate at what gas densities the \ce{SO_2}, \ce{CH_3OH}, and \ce{H_2CO} emission thermalises. We have run the calculations for each molecular species, including all transitions with frequencies $<$1000 GHz, but limiting the transitions to those with $E_\textnormal{up}\leq$300 K and $A_\textnormal{ul}\geq$10$^{-6}$ s$^{-1}$ to cover similar properties as our observed transitions, as only using the detected transitions did not yield sensible results. The number of transitions is too sparse to derive meaningful rotational temperatures for the entire grid. We ran the calculations separately for the A- and E-transitions of \ce{CH_3OH} and for the ortho- and para-transitions of \ce{H_2CO}. The calculations are run over grids of $T_\textnormal{kin}$ (10 K to 200 K in steps of 10 K) and $\log_{10}\left(n_\textnormal{\ce{H_2}}\right)$ (5 to 12 in steps of 0.1, with $n_\textnormal{\ce{H_2}}$ in units of cm$^{-3}$). In our calculations, we have kept the column density at a fixed value of $N$=10$^{12}$ cm$^{-2}$, which ensures optically thin emission and we used, similarly as in Section \ref{sec:Res-H2CO}, the default line width of $\sim$1 km~s$^{-1}$. For each calculation, we infer the rotational temperature by fitting a straight line ($y=ax+b$) through all the retrieved RADEX integrated fluxes, where the rotation temperature is given by the slope, $a=-1/T_\textnormal{rot}$. Figure \ref{fig:RADEX} displays the percentile difference between the input kinetic temperature and the derived rotational temperature ($\Delta T$) from the RADEX calculations. For smaller percentages, the lines are thermalised. The contours (solid, dashed, and dashed-dotted) indicate where the differences are, respectively, $\Delta T$=1\%, 5\%, and 10\%. Above these contours (i.e. for higher densities), the difference becomes smaller and the emission is thermalised, whereas below these contours the differences become larger and the lines are not thermalised. Figure \ref{fig:RADEX} also indicates the gas density, up to a height of 20 au, at the location of the IRS~48 dust trap, following the DALI model of \citet{LeemkerEA23} (see also Figure \ref{fig:ModelSetup}), and the derived excitation conditions from the rotational diagrams: $T\sim$55 K for \ce{SO_2}, $T\sim$130 K for \ce{CH_3OH}, and the lower end of the derived temperature range ($T\sim$150 K) for \ce{H_2CO}. \\
\indent We find that \ce{SO_2} thermalises at densities of $n_\textnormal{\ce{H_2}}\sim$10$^{8}$ cm$^{-3}$ and higher. As the rotational temperature of \ce{SO_2} is well-constrained at $T\sim$55 K, this requires that the \ce{SO_2} emission should be thermalised. However, densities of $n_\textnormal{\ce{H_2}}\sim$10$^8$ cm$^{-3}$ are only achieved near the midplane of the disk at heights of $\sim$5 au ($z/r<$0.1, see also Figure \ref{fig:ModelSetup}), suggesting that the \ce{SO_2} may originate from deep inside the disk. The emitting height of \ce{SO_2} is further discussed in Section \ref{sec:TempStruc}. \\
\begin{figure*}[h!]
    \centering
    \includegraphics[width=\textwidth]{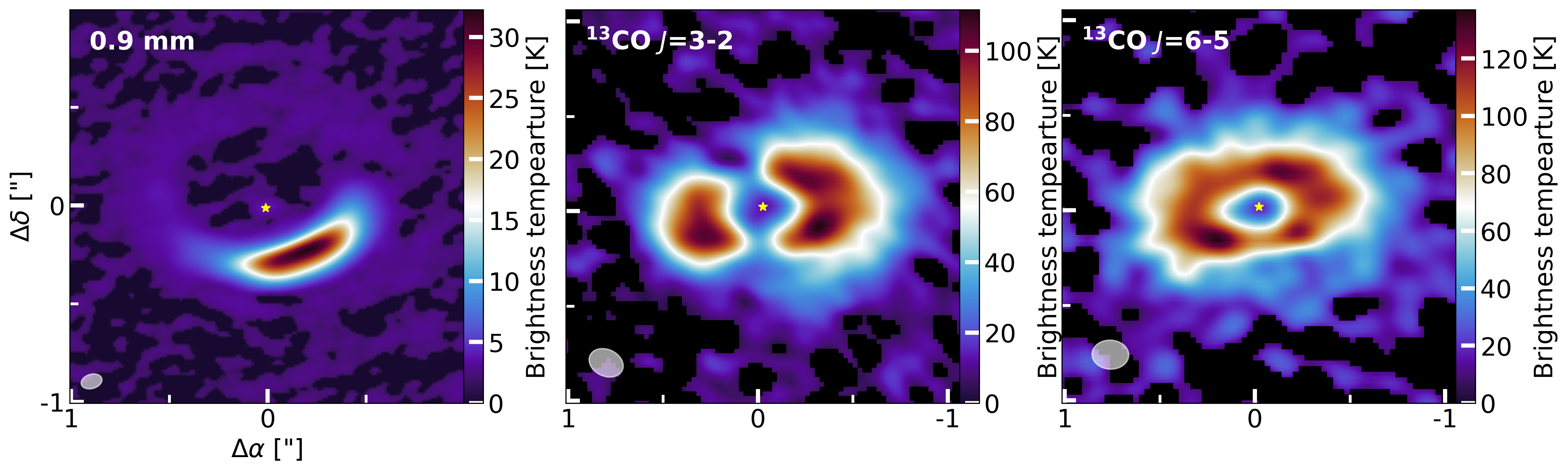}
    \caption{Brightness temperature maps of the dust continuum (left) and the \ce{^{13}CO} $J$=3-2 (middle panel) and \ce{^{13}CO} $J$=6-5 (right panel) transitions. The yellow stars in the centre indicate the approximate location of the host star, while the resolving beams are indicated by the white ellipses.}
    \label{fig:BT-Maps}
\end{figure*}
\indent For \ce{H_2CO} we find that the emission thermalises at similar densities as \ce{SO_2}, however, the derived temperatures are significantly higher, suggesting that the \ce{H_2CO} emission does not originate from the same layer as \ce{SO_2}. \ce{CH_3OH} requires, on the other hand, even higher densities of $n_\textnormal{\ce{H_2}}>$10$^9$ cm$^{-3}$. Combining these dust trap densities with the derived rotational temperatures, we infer that the emission of both \ce{CH_3OH} and \ce{H_2CO} may be sub-thermally excited. Therefore, the presented rotational temperatures are considered to underestimate the kinetic temperature and non-LTE approaches should be used to correctly derive the excitation conditions. As \ce{CH_3OH} requires higher gas densities to be thermalised compared to \ce{H_2CO} (i.e. the differences, $\Delta T$, are larger), we expect that the underestimation of the kinetic temperature is larger for \ce{CH_3OH} than for \ce{H_2CO} (as indicated in Figure \ref{fig:RADEX}). Therefore, we expect that the \ce{CH_3OH} and \ce{H_2CO} may actually probe very similar temperatures and, as suggested by \citet{vdMarelEA21}, that the species may originate from the same high layer inside the disk. \\
\indent We reiterate that the IRS~48 disk has exceptionally low gas and dust masses. For disks with higher masses and, therefore, higher densities, the molecular emission is more likely to be thermally excited. Therefore, observations (or surveys) of massive disks, covering multiple ($\geq$3 or at least 2, if the upper level energies are significantly different) transitions of the same molecular species, could be used to infer the (gas) temperature and chemical structure from their molecular line emission. For these less massive disks, a non-LTE analysis is a necessary and powerful tool to infer the excitation conditions. 

\subsection{Temperature and vertical structure} \label{sec:TempStruc}
In the following section, we discuss the resulting temperature (Section \ref{sec:TS}) and vertical structure (Section \ref{sec:VS}) of the disk, as interpreted from our results.

\subsubsection{Temperature structure} \label{sec:TS}
Using our temperature maps (top row in Figure \ref{fig:Maps}), we assess how the temperature may vary across the IRS~48 dust trap. In Figure \ref{fig:TUncertainty}, we display the derived lower and upper uncertainties for the temperature maps of \ce{SO_2} and \ce{CH_3OH}. The uncertainties are found to be the highest for pixels near the edge, for which often also the highest temperatures are found. This is very likely due to the fact that fewer transitions are detected and the line emission is not as strong as for pixels closer to the centre (see also Figure \ref{fig:NoDetections}). The temperature map for \ce{CH_3OH} does not show a clear temperature structure but is rather constant across the entire disk with temperatures of $T\gtrsim$125 K. The \ce{SO_2} map, on the other hand, shows hints of a potential radial temperature gradient, which is generally expected to be present in disks. The temperature closer to the host star is hotter ($T\sim$70 K) compared to the temperature at the outer edge ($T\sim$40 K). \\
\indent Overall, we find that the uncertainties for \ce{SO_2} and \ce{CH_3OH} are nearly uniform ($\sigma_T\lesssim$5 K) across the molecular emission extent. Additionally, Figure \ref{fig:NUncertainty} show the uncertainties for the column density maps. We find that the uncertainties on the column density are uniform across the maps of all three molecular species, on the order of $\lesssim$10\%. As our temperature uncertainties are on the order of $\sigma_T\lesssim$5 K, we expect this radial temperature gradient in the \ce{SO_2} map to be real. The observed radial decrease in the temperature is on the order of 10 K to 30 K when considering the uncertainties. The DALI model suggests, on the other hand, that the temperature for the expected emitting layer of \ce{SO_2} (see Section \ref{sec:VS}), remains fairly constant between 40 and 70 K. \\
\begin{figure}[ht!]
    \centering
    \includegraphics[width=\columnwidth]{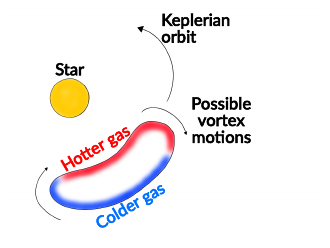}
    \caption{Cartoon visualising the expected effect of the vortex on the temperature structure. Hotter gas, due to UV heating, will be counteracted by colder gas being brought in due to the potential vortex motions, which primarily acts in the midplane. This effect is not observed in the IRS~48 disk, unless the temperatures variations are very small.}
    \label{fig:Cartoon-Temp}
\end{figure}
\indent As the asymmetric dust trap of IRS~48 may be the result of a vortex present in the disk, this vortex could also influence the observable temperature structure. As discussed by \citet{LR14}, if the vortex is the result of the RWI, the vortex must be anticyclonic (i.e. rotate clockwise) for it to generate a pressure maximum that can trap the dust. Most of the recent work on vortices has gone into their existence, evolution, and stability (e.g. \citealt{LovelaceEA99,MeheutEA12,MeheutEA12b,LH13,MeheutEA13}), or into their relationship with forming planets (e.g. \citealt{HammerEA17,HammerEA19,HammerEA21,HammerEA23}). Little work (see, for example, \citealt{HuangEA18,LeemkerEA23}) has focused on how the gas and chemistry may be affected by the vortices. Vortices perturb the motion of the gas, so intuitively, such a perturbation may also be visible in the temperature structure. If the gas perturbation follows the anticyclonic rotation of the vortex, gas with slightly higher temperature would be rotating away from the host star, whereas cooler gas would move towards the host star. This is also visualised in Figure \ref{fig:Cartoon-Temp}. The strength of this effect would depend on the rotational timescale of the vortex, which itself is set by the vorticity \citep{MeheutEA13}: the faster the vortex rotates, the stronger the effect will be. As vortices predominantly arise in the midplane, their effect on the dust and gas should be most notable for molecular emission originating from close to the midplane, but may be negligible in elevated layers, where UV heating dominates. As we expect \ce{SO_2} to emit from close to the midplane and we do not see an azimuthal variation that could be induced by a potential vortex, the temperature variation induced by such a vortex can only be on the order of $\sim$2 K, given the small uncertainties. \ce{CH_3OH} is, on the other hand, together with \ce{H_2CO} expected to emit from an elevated layer $\sim$10-15 au ($z/r\sim$0.17-0.25) above the midplane \citep{vdMarelEA21}. Therefore, we expect not to see any variation in the temperature related to a potential vortex. Only if vertical mixing, which is expected to play a role in liberating the molecular species from the icy dust mantles in the IRS~48 disk, can transmit the effect from a vortex to elevated regions, a different temperature structure may be found in the molecular emitting layers. Detailed models, that constrain the vorticity and include vertical mixing, can be used to investigate the interplay between the dust and gas in vortices and, subsequently, are necessary to investigate the potential influences of vortices on the temperature structures of disks. 

\subsubsection{Vertical structure} \label{sec:VS}
\indent In addition to the temperature maps, we have measured the temperature using the brightness temperature ($T_\textnormal{b}$) of optically thick emission, i.e. from the dust continuum or from one of the \ce{CO} isotopologues. Using Planck's law, we have calculated the brightness temperature of the high-resolution, optically thick 0.9 mm continuum emission \citep{YangEA23} and of the \ce{^{13}CO} $J$=3-2 \citep{LeemkerEA23} and $J$=6-5 \citep{vdMarelEA16} transitions. The brightness temperature maps are displayed in Figure \ref{fig:BT-Maps}. We find a peak brightness temperature of $T_\textnormal{b}\sim$32 K for the dust continuum, which is only slightly higher ($\Delta T_b\sim$5 K) than that found in \citet{vdMarelEA21}, likely due to the difference in the beam sizes. The higher brightness temperature determined here confirms the notion that the temperature of the dust in the disk's midplane is too high for \ce{CO} freeze-out and in situ COM formation through the hydrogenation of \ce{CO} \citep{WK02,FuchsEA09,SantosEA22}. However, \ce{CO} may still react with solid-state \ce{OH} to form \ce{CO_2}, even though the temperatures are above its sublimation temperature \citep{TvSEA22}. \\
\indent As the \ce{CO} emission is affected by foreground cloud absorption \citep{BrudererEA14}, the temperature can only be estimated for select emission channels. We find peak brightness temperatures of $T_\textnormal{b,3-2}\sim$110 K and $T_\textnormal{b,6-5}\sim$135 K, from the absorption free channels of the optically thick \ce{^{13}CO} $J$=3-2 and $J$=6-5 transitions, respectively. Based on these derived temperatures, and those derived for \ce{SO_2}, \ce{CH_3OH} and \ce{H_2CO}, we can infer the vertical emission structure of the disk. We expect \ce{CH_3OH} and \ce{H_2CO} to trace the highest temperatures, given their derived temperatures and the notion that the emission is sub-thermally excited, and, therefore, they should arise from a higher emitting layer. The \ce{^{13}CO}, probing a slightly colder temperature, should originate from a slightly deeper layer in the disk, under the assumption that the emission is fully resolved \citep{LeemkerEA22}, while the \ce{SO_2} should originate from the deepest vertical layer. The notion that \ce{SO_2} originates from deep inside the disk, based on the temperature, is supported by our RADEX calculations (see Section \ref{sec:STE}), which show that \ce{SO_2} becomes thermalised at densities of $n_\textnormal{gas}\sim10^8$ cm$^{-3}$. These densities are only reached in the layers close to the midplane, with heights of $\lesssim$5 au (or z/r$<$0.1, see also Figure \ref{fig:ModelSetup}). Therefore, under the assumption of thermalised emission, the \ce{SO_2} emission must originate from this layer just above the midplane. We also consider the radial locations of the emission to support our vertical structure posed above. We note that the \ce{SO_2} emission peaks at larger distances, by $\sim$10 au, than both \ce{CH_3OH} and \ce{H_2CO} (see \citealt{BoothEA24}).

\subsection{Optical depth and emission structure} \label{sec:ODES}
Below, we discuss the optical depth of the various molecules (Section \ref{sec:OD}). In addition, we also discuss the required radial extent of the emission to become optically thick.

\subsubsection{Optical depth} \label{sec:OD}
\begin{figure}[ht!]
    \centering
    \includegraphics[width=\columnwidth]{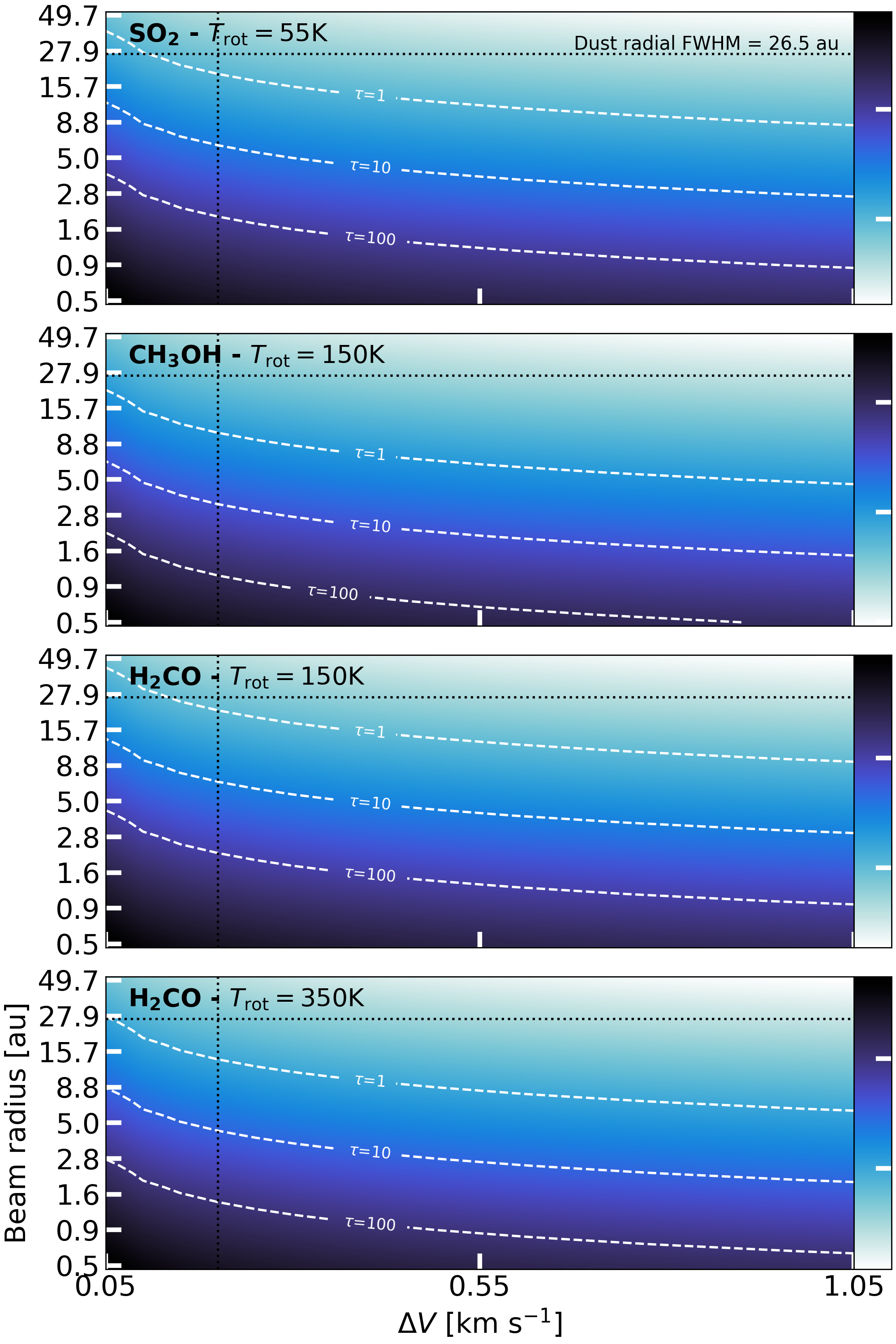}
    \caption{Optical depth (given in $\log_{10}$-space) grids for a range of beam radii (assuming circular beams) and turbulent line widths ($\Delta V$) for \ce{SO_2} $J$=6$_{4,2}$-6$_{3,3}$ (at $T_\textnormal{rot}$=55 K), \ce{CH_3OH} $J$=13$_{-1}$-13$_0$ (at $T_\textnormal{rot}$=150 K, accounting for the underestimation of the temperature due to sub-thermal excitation), and \ce{H_2CO} $J$=5$_{1,5}$-4$_{1,4}$ (at $T_\textnormal{rot}$=150 K and $T_\textnormal{rot}$=350 K). The contour lines indicate values of $\tau$=1, 10, and 100. The dotted black line indicates the radial FWHM of the dust trap ($FWHM_\textnormal{dust}\simeq$26.5 au), found using the high-resolution observations of \citet{YangEA23}. The vertical dashed line indicates a turbulent line width of $\Delta V_\textnormal{line}$=0.2 km~s$^{-1}$, the lower value used in our analysis.}
    \label{fig:ODGrids}
\end{figure}

\begin{figure}[ht!]
    \centering
    \includegraphics[width=\columnwidth]{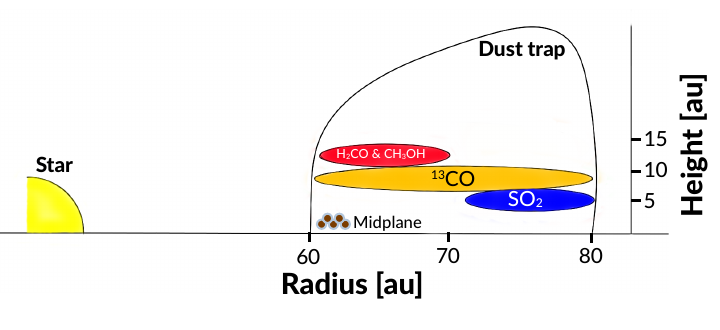}
    \caption{Cartoon visualising our proposed radial vertical emission scheme, containing \ce{SO_2} (blue), \ce{^{13}CO} (orange), and \ce{H_2CO} and \ce{CH_3OH} (red). The radial and vertical emission locations are based on the derived temperatures, the proposed ``sliver''-theory, and the assumption that the \ce{^{13}CO} emission is spatially resolved.}
    \label{fig:Cartoon}
\end{figure}

For the chosen turbulent line width ($\Delta V_\textnormal{line}$=0.2 km~s$^{-1}$) our analysis points towards potentially optically thin emission. However, it is still possible that the molecular emission, in particular that of \ce{H_2CO}, is optically thick. Considering a previously derived isotopologue ratio, \ce{H_2CO}/\ce{H_2 ^{13}CO}$\sim$4 \citep{BoothEA24}, it is very likely that its emission is optically thick. As the spatial resolution of the used observations is low, one likely explanation for our analysis pointing towards optically thin emission is that the actual emitting areas are smaller than currently can be probed. As can be seen in Figure \ref{fig:M0s}, the radial extent is of a similar size as that of the beam. The azimuthal extent of both \ce{SO_2} and \ce{H_2CO} is, on the other hand, larger than a single beam. Therefore, we conclude that the azimuthal extent is resolved, whereas the radial one is not. \\
\indent To test for what emitting areas, parameterised by the beam radius (assuming a circular beam), and values for the local (turbulent) line widths we can ensure optically thick emission, we have explored grids over both parameters. We have run the grids for the same transitions as used throughout our analysis: \ce{SO_2} $J$=6$_{4,2}$-6$_{3,3}$, \ce{CH_3OH} $J$=13$_{-1}$-13$_0$, and \ce{H_2CO} $J=$5$_{1,5}$-4$_{1,4}$. For \ce{H_2CO}, we have used a higher resolution (both spatial and spectral) transition that has been presented in \citet{BoothEA24} (see also Figure \ref{fig:M0s-H2CO}), which is of the same spectral and spatial resolution as the \ce{SO_2} and \ce{CH_3OH} transitions. In our calculations of the optical depth, we have used fixed rotational temperatures of $T_\textnormal{rot}$=55 K for \ce{SO_2}, $T_\textnormal{rot}$=150 K for \ce{CH_3OH}, and $T_\textnormal{rot}$=150 K and $T_\textnormal{rot}$=350 K for \ce{H_2CO} to represent the range of temperatures found. Our grid of the line width extends up to 1.0 km~s$^{-1}$. If the line profiles are only thermally broadened, they will be on the order of $\sim$0.15 km~s$^{-1}$, but local turbulence can additionally broaden the lines by $\sim$0.1 km~s$^{-1}$ (see, e.g., \citealt{FlahertyEA20,FlahertyEA24,PanequeEA24}). \\ 
\indent The inferred optical depths for our grids of turbulent line widths and beam radii are displayed in Figure \ref{fig:ODGrids}. As can be seen, we require beam radii of $\lesssim$8 au for the \ce{H_2CO} emission to become optically thick at a temperature of $T_\textnormal{rot}\sim$350 K. For the lower temperature of $T_\textnormal{rot}\sim$150 K, a larger beam area of $\lesssim$16 au is sufficient for optically thick emission. These lower temperatures should hold for the outer edges of the dust trap, whereas the warmer temperature accounts for the central region (see also Section \ref{sec:Res-H2CO}). The \ce{SO_2} emission also becomes optically thick for an emitting radius of $\lesssim$8 au, whereas the \ce{CH_3OH} emission requires slightly smaller radii. For smaller line widths ($\Delta V_\textnormal{line}<$0.2 km~s$^{-1}$), we find that the emission becomes optically thick for larger beam areas. \\
\indent We, thus, propose a scenario in which the emission (in particular that of \ce{H_2CO}) originates from a small radial ``sliver'' with an approximate width of $\lesssim$10 au. This ``sliver'' must be located at the inner edge of the dust trap for \ce{CH_3OH} and \ce{H_2CO} to conform with the current hypothesis that the observed molecules are in the gas phase following thermal sublimation (see, e.g., \citealt{vdMarelEA21,BoothEA21,BoothEA24}). While \ce{SO_2} emission is known to originate from a larger radial distance \citep{BoothEA24}, it is still possible that the emission originates from a similar ``sliver'', only placed at a larger radial distance, coming from close to the midplane, and is located closer to the outer edge of the dust trap. \\
\indent For comparison, we have estimated, from the high-resolution dust observations \citep{YangEA23}, that the dust trap has a width of $\sim$26.5 au (see also Figure \ref{fig:ODGrids}). Interestingly, from imaging of the \ce{H_2CO} $J$=5$_{1,5}$-4$_{1,4}$ transition with superuniform weighting (see right panel of Figure \ref{fig:M0s-H2CO}), a smaller width of the emission is inferred. We note that the radial extent is still unresolved with this weighting scheme. The resulting beam (0.22"$\times$0.18") has a width of $\sim$25 au, which is bigger than the proposed width of the ``sliver''. Finally, we note that one of the models presented by \citet{vdMarelEA21} also suggests this sliver theory for the \ce{H_2CO} theory. In particular, their full model, where the large grain population is increased throughout the entire dust column instead of settled to the midplane, with a surface density of $\Sigma_\textnormal{d}$=5.0 g~cm$^{-2}$ at 60 au, has the \ce{H_2CO} distributed over the entire vertical extent, but radially only between 60 and 70 au (see their Figure C.2). We have visualised our proposed emission scheme in Figure \ref{fig:Cartoon}, based on the temperatures as discussed in Section \ref{sec:TempStruc} and the potential radial extent of the molecules as discussed here. \\
\indent The discussion above suggests that the emitting areas of the molecules are radially unresolved and, subsequently, the emission is beam diluted. This has previously already been discussed in \citet{BrunkenEA22} and \citet{BoothEA24}. Beam dilution results in an underestimation of the inferred column densities. Higher resolution observations with spatial resolutions of $<$0.1" (translating to a resolution of $<$13.5 au for the distance of IRS~48) are needed to obtain better constraints on the column densities. Using the following equation,
\begin{align*}
    \textnormal{DF}^{-1} = \frac{\Omega_\textnormal{source}^2}{\Omega_\textnormal{beam}^2 + \Omega_\textnormal{source}^2}
\end{align*}
we provide an estimate on the dilution factor for the molecular emission, assuming the emission comes from a circular region with a total width of 10 au, as proposed above with our ``sliver'' theory. As determined by the column densities on a pixel-by-pixel basis, we take the $\Omega_\textnormal{beam}^2$ to be equal to the beam size of the current observations, whereas we set $\Omega_\textnormal{source}^2$ equal to the area of the circular region. This yields $\textnormal{DF}^{-1}\sim$0.002 or, in other words, an underestimation of the column density by a factor of $\sim$500.  \\
\indent For our derived column densities (see Section \ref{sec:Results}) and given the uncertain optical depths, we derive lower limits on the fractional abundances, with respect to \ce{H_2} ($\Sigma_g$=0.25 g~cm$^{-2}$ or $N_\textnormal{\ce{H_2}}\sim$7$\times$10$^{22}$ cm$^{-3}$; \citealt{vdMarelEA16,LeemkerEA23}), of $N_\textnormal{\ce{SO_2}}/N_\textnormal{\ce{H_2}}\sim$7$\times$10$^{-10}$-3$\times$10$^{-9}$ and $N_\textnormal{\ce{CH_3OH}}/N_\textnormal{\ce{H_2}}\sim$(1-8)$\times$10$^{-9}$. Assuming that we underestimate the column densities by a factor of $\sim$500, these fractional abundances become $N_\textnormal{\ce{SO_2}}/N_\textnormal{\ce{H_2}}\sim$3$\times$10$^{-7}$-2$\times$10$^{-6}$ and $N_\textnormal{\ce{CH_3OH}}/N_\textnormal{\ce{H_2}}\sim$7$\times$10$^{-7}$-4$\times$10$^{-6}$. Compared to two other Herbig stars, HD~100546 and HD~169142, we find that our derived column densities for \ce{CH_3OH} are of the same order of magnitude (\citealt{BoothEA23} and Evans et al. in prep.). However, we note that our column densities may be underestimated by this factor of $\sim$500.

\subsubsection{Continuum affecting the line emission} \label{sec:Continuum}
One additional concern for our derived excitation properties is the potential effect of the dust continuum on the observed molecular emission, in particular, that of \ce{CH_3OH} and \ce{H_2CO}. The continuum emission can impact the line emission in two ways \citep{BoehlerEA17,WeaverEA18,IsellaEA18,RabEA20,BosmanEA21,RosottiEA21,NazariEA23}: (1) if the lines themselves are optically thick, the continuum subtraction procedure may overestimate the contribution of the continuum at the line centre. The subtraction would, subsequently, result in an oversubtraction of the continuum. This effect maximises once the continuum also becomes optically thick. (2) On the other hand, if the line is optically thin and the continuum is optically thick, up to half of the total emission, originating from the backside of the disk, may be blocked by the continuum. \\
\indent As it is expected that the \ce{H_2CO} emission is optically thick and polarisation observations of the IRS~48 continuum and accompanying modelling works by \citet{OhashiEA20} suggest that the dust continuum at 0.86 mm ($\sim$348.6 GHz) is optically thick ($\tau_\textnormal{dust}\sim$7.3), it is not immediately clear which effect is at play. Nonetheless, the channel maps of the \ce{CH_3OH} $J$=13$_{-1}$-13$_0$ and \ce{H_2CO} $J$=5$_{1,5}$-4$_{1,4}$ transitions show that the observed emission is affected by the continuum, see Figure \ref{fig:ChannelMaps}. The channel maps for \ce{CH_3OH} and \ce{H_2CO} show a decrease, most notable in the 4.8-6.6 km~s$^{-1}$ velocity range for \ce{CH_3OH} and 3.9-7.5 km~s$^{-1}$ range for \ce{H_2CO}, in the line emission near the peak location of the dust continuum, which is not visible in those of the \ce{SO_2} $J$=6$_{4,2}$-6$_{3,3}$ transition. Therefore, we expect that we are not probing the full line flux of these molecules, either due to continuum oversubtraction or the continuum blocking emission originating from the backside of the disk. Given that the effect is most notable in the channel maps of \ce{CH_3OH} and \ce{H_2CO}, and not in those of \ce{SO_2}, we propose that continuum oversubtraction is the likely effect. Especially considering that the \ce{SO_2} originates from near the midplane, where the secondary effect (dust blocking the emission) could be even stronger than for the elevated layers, i.e. more than half of the emission could be blocked. This suggests that \ce{CH_3OH}, in addition to \ce{H_2CO}, should be optically thick. Therefore, we propose that the ``sliver" theory mentioned above also should hold for \ce{CH_3OH}. This, in addition to the beam dilution, results in an underestimation of the column density by as much as a factor of 500. 

\subsection{Future observations} \label{sec:FutureObs}
Any conclusions on the chemistry are currently limited by the spatial and spectral resolution of the observations. Therefore, to be able to infer more information on the chemistry (i.e. column densities), we need higher resolution (both spatially and spectrally) observations that are able to test our ``sliver'' theory proposed in Section \ref{sec:ODES} for \ce{H_2CO} and potentially \ce{CH_3OH}. At the distance of IRS~48 ($d\sim$135 pc), an angular resolution of $\sim$0.1" is needed to retrieve the necessary beam radius of $\sim$8 au. In addition, these observations may provide more insights into the varying temperature structure, potentially related to the influence of a vortex. \\
\indent Furthermore, a spectral resolution of $\sim$0.1 km~s$^{-1}$ will ensure resolved line profiles for line widths of $\gtrsim$0.5 km~s$^{-1}$. This spectral resolution will allow one to infer the vertical emission heights from the channel maps, as recently has been done for a multitude of other inclined disks \citep{LawEA21,PanequeEA23,LawEA23,LawEA23b,LawEA24,UrbinaEA24}. A vertical decomposition of the emission layers could also confirm our notion that the \ce{CH_3OH} and the \ce{H_2CO} must come from higher emitting layers compared to \ce{^{13}CO}, while the \ce{SO_2} must originate from deep inside the disk. Additionally, the proposed resolutions will be sufficient to study kinematical signatures that deviate from Keplerian emission at the location of the dust trap, as was performed for the asymmetric HD~142527 disk using CO isotopologues \citep{BoehlerEA21}. \\
\indent Finally, as seen in Figure \ref{fig:RatioMaps}, many of the line ratios are limited by the temperature sensitivity of the transitions. In order to better constrain the temperature through these ratios, transitions with higher upper level energies need to be observed. These lines, in addition to the higher resolution observations, will ensure proper constraints on the temperature, not influenced by effects such as beam dilution.

\section{Timescales: vertical mixing, desorption, and photodissociation} \label{sec:Timescales}
\begin{figure*}[ht!]
    \centering
    \includegraphics[width=\textwidth]{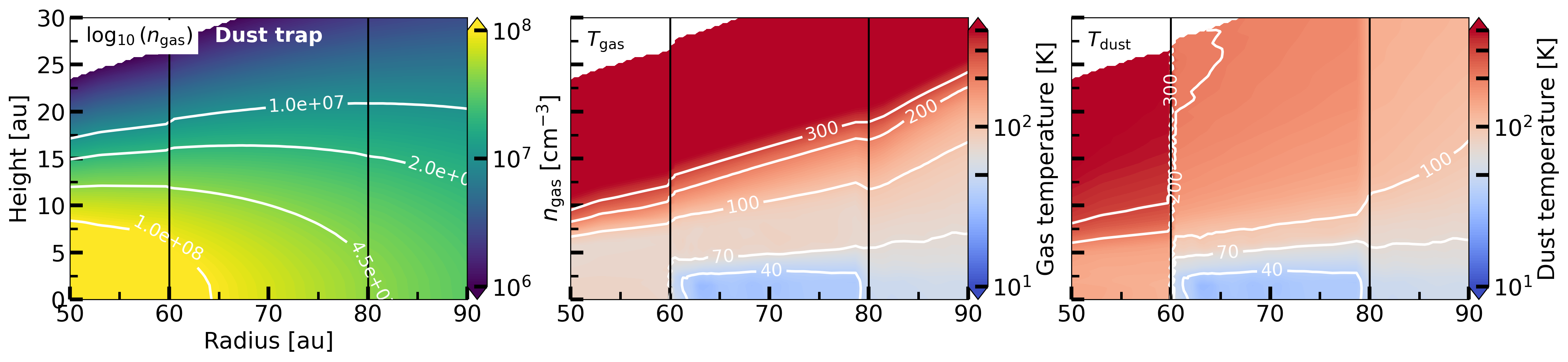}
    \caption{Zoom-in of the model structure of the dust trap region of the IRS~48 disk of the gas density ($n_\textnormal{gas}$ in cm$^{-3}$, left panel) and gas and dust temperature (both in K, left and right panels, respectively). The vertical bars indicate the location of the dust trap, extending from 60 to 80 au, while the white contours in the left most panel indicate gas densities between 10$^7$ and 10$^8$ cm$^{-3}$. The images have been adapted from \citet{LeemkerEA23}.}
    \label{fig:ModelSetup}
\end{figure*}
\subsection{Methodology}
To further determine what sets the observable molecular abundance in the IRS~48 disk, we have calculated multiple timescales at the location of the dust trap: the vertical turbulence mixing timescale, the desorption and freeze-out timescales of \ce{SO_2}, \ce{CH_3OH}, and \ce{H_2CO}, and their photodissociation timescales. Interestingly, the disk of IRS~48 has very low gas and dust masses, with a gas mass on the order of $M_\textnormal{gas}\sim$1.4$\times$10$^{-4}$ M$_\odot$ \citep{BrudererEA14}, obtained through modelling the observed \ce{C^{17}O} $J$=6-5 flux and found to be consistent with the mass derived from the \ce{C^{18}O} observations \citep{vdMarelEA16}, and a dust mass of $M_\textnormal{dust}\sim$(6.2$\pm$0.6)$\times$10$^{-5}$ M$_\odot$ \citep{TemminkEA23}. These masses are lower compared to other Herbig stars, which have disk masses on the order of $\sim$10$^{-2}$ M$_\odot$ (e.g. \citealt{StapperEA24}). Therefore, the gas densities in the IRS~48 dust trap, even near the midplane, are only of the order of 10$^{8}$ cm$^{-3}$ and drop to 10$^{6}$ cm$^{-3}$ in the surface layers.  We have used the inferred properties (i.e. $n_\textnormal{gas}$, $T_\textnormal{gas}$, and $T_\textnormal{dust}$) of the IRS~48 disk as modelled by \citet{LeemkerEA23} using the thermochemical code DALI \citep{BrudererEA12,BrudererEA13}. This DALI model is built upon an existing model from \citet{BrudererEA14}, \citet{vdMarelEA16}, and \citet{vdMarelEA21}, and is described by a radial surface density profile that uses the self-similar solution for a viscously evolving disks \citep{LP74,HartmannEA98}. It consists of deep dust and gas cavities between 1 and 60 au, while the dust trap is placed at 60 to 80 au. The model is separated into two different models: one representing the dust-trap side (where the dust density distribution has been altered under the assumption that the small grains have grown to larger sizes) and one tailored to the other side of the disk. This model is able to reproduce the continuum emission (at 0.44 mm and 0.89 mm) and the emission of various \ce{CO} isotopologues. The gas temperature is also well constrained within the model, as it is able to reproduce emission of the \ce{CO} $J$=6-5 transition. We refer the reader to \citet{LeemkerEA23} for all details on the model (see their Appendix B). Figure \ref{fig:ModelSetup} displays the gas density (in cm$^{-3}$) and the gas and dust temperature of the DALI model in the region of the dust trap. Below, we discuss how the timescales are calculated, whereas their implications on the emission are further discussed in Section \ref{ref:TS-Disc}. \\ 
\indent We determine the turbulent vertical mixing timescale following the description provided by \citet{XieEA95} and \citet{SemenovEA06}, where the vertical mixing is set by the turbulent diffusion. The mixing timescale can be approximated as $t_\textnormal{mix}\sim H^2/D_\textnormal{turb}$, where $H$ is the scale height (taken as $H=c_\textnormal{s}/\Omega_\textnormal{K}$) and $D_\textnormal{turb}=\alpha c_\textnormal{s}^2/\Omega_\textnormal{K}$ is the turbulent diffusion coefficient (in cm$^{-2}$~s$^{-1}$). Here, $c_\textnormal{s}$ and $\Omega_\textnormal{K}$ denote the sound speed (taken to be isothermal) and the Keplerian angular frequency, respectively, while $\alpha$ parametrises the viscosity strength. We have determined the vertical mixing timescales for three commonly used values of $\alpha$: 10$^{-4}$, 10$^{-3}$, 10$^{-2}$. For comparison, \citet{YangEA24} find turbulent $\alpha$ parameters, using the scale height ratio prescription of \citet{YL07} and assuming a grain size of 140 $\mathrm{\mu}$m and a dust surface density of $\Sigma_\textnormal{d}$=1 g~cm$^{-2}$, ranging from 10$^{-4}$ to 5$\times$10$^{-3}$ for gas-to-dust mass ratios of 3-100 in the IRS~48 disk. As the gas-to-dust mass ratio is expected to be much lower in the dust trap $\sim$0.009-0.04 \citep{OhashiEA20,vdMarelEA16,LeemkerEA23}, the $\alpha$ parameter could be even higher. Therefore, our range of $\alpha$ values encapsulates the expected range well. We note that a higher $\alpha$ parameter, for example $\alpha$=0.1, only changes our results by a factor of 10, i.e. the turbulent diffusion coefficient would increase by this factor, whereas the mixing timescale would decrease by this factor. \\
\indent The desorption ($k_\textnormal{d}$) and freeze-out ($k_\textnormal{f}$) rates (in s$^{-1}$) are determined for each molecular species $i$ following the equations given in \citet{TA87} (see also \citealt{WalshEA10} and \citealt{LeemkerEA21}):
\begin{align}
    k_\textnormal{d,i} & = v_0(i)\exp\left(-\frac{E_{\textnormal{b},i}}{k_\textnormal{B}T_\textnormal{dust}}\right) \nonumber \\ 
                       & = \sqrt{\frac{2n_\textnormal{s}E_{\textnormal{b},i}}{\pi^2m_i}}\exp\left(-\frac{E_{\textnormal{b},i}}{k_\textnormal{B}T_\textnormal{dust}}\right) \label{eq:Desorp}\\
    k_\textnormal{f,i} & = \sqrt{\frac{k_\textnormal{B}T_\textnormal{gas}}{m_i}}\pi a_\textnormal{grain}^2n(\textnormal{grain})S. \label{eq:Freeze-out}
\end{align}
Here, $T_\textnormal{gas}$ and $T_\textnormal{dust}$ represent, respectively, the gas and dust temperature. $E_{\textnormal{b},i}$ denotes the binding energy (in K), while $m_i$ is molecular mass. $v_0(i)$ is the characteristic vibrational frequency of species $i$, which has been approximated with the harmonic oscillator relation \citep{HasegawaEA92}. $n_\textnormal{s}$ is the number density of surface sites ($n_\textnormal{s}$=1.5$\times$10$^{15}$ cm$^{-2}$) and $n(\textnormal{grain})$ is the grain number density of the DALI model. Additionally, $a_\textnormal{grain}$ is the grain size, which we have taken to be the average grain size in the DALI model (see Equation 14 of \citealt{FacchiniEA17}), and $S$ is the sticking coefficient, assumed to be unity for temperatures below the desorption temperature. For our molecular species, we have taken binding energies of $E_\textnormal{b,\ce{SO_2}}$=3254 K \citep{PerreroEA22}, $E_\textnormal{b,\ce{CH_3OH}}$=4930 K \citep{BB07}, and $E_\textnormal{b,\ce{H_2CO}}$=3260 K \citep{NobleEA12,PenteadoEA17}. We note that generally both the freeze-out and desorption of a molecular species cannot be described by a single binding energy, but that a range of values is more realistic, as the binding energy may depend, for example, on the composition of the ice (see, e.g., \citealt{MinissaleEA22,Furuya24}). However, we use the listed values to obtain the approximate timescales for our target species, as we expect these values to provide reasonable timescales with respect to one another. The timescales for both desorption and freeze-out were estimated by taking the inverse of their respective rates. \\
\indent Finally, the photodissociation timescales are determined using the following equation,
\begin{align}
    t_\textnormal{pd} = \frac{1}{kG_0}.
\end{align}
$k$ is the photodissociation rate, which is taken to be the rate for a blackbody with a temperature of $T=$10,000 K when scaled to have the same integrated intensity between 912 and 2000 \r{A} as the interstellar radiation field (IRSF) and $G_0$ is the integrated intensity in the FUV band, normalised to the Draine field, as calculated within DALI (see, e.g., Figures C.1 and C.2 in \citealt{vdMarelEA21}). We have used the following values for $k$, $k_\textnormal{\ce{SO_2}}$=1.6$\times$10$^{-9}$ s$^{-1}$, $k_\textnormal{\ce{CH_3OH}}$=5.3$\times$10$^{-10}$ s$^{-1}$, $k_\textnormal{\ce{H_2CO}}$=9.6$\times$10$^{-10}$ s$^{-1}$ \citep{HeaysEA17}.

\subsection{Timescale comparisons} \label{ref:TS-Disc}
The determined timescales are shown in Figures \ref{fig:Model-Diffusion}, \ref{fig:Model-FD}, and \ref{fig:Model-PD}, respectively. As shown in these figures, we find (for all three molecules, \ce{SO_2}, \ce{CH_3OH}, and \ce{H_2CO}) $t_\textnormal{desorption} < t_\textnormal{photodissociation} < t_\textnormal{mixing}$ in the higher emitting layers, whereas we see $t_\textnormal{photodissociation} < t_\textnormal{mixing} < t_\textnormal{desorption}$ in the midplane. \\
\indent As can be seen in the top panels of \ref{fig:Model-FD}, desorption occurs nearly instantaneous for emitting heights of $>$5 au ($z/r>$0.08) for both \ce{SO_2} and \ce{H_2CO}, where the temperature is higher than the freeze-out temperature. As the binding energy of \ce{CH_3OH} is higher, these ices will desorp at higher temperatures (given the $T$ in the exponential in Equation \ref{eq:Desorp}) and, therefore, at larger heights in the disk ($\gtrsim$10 au, or $z/r\gtrsim$0.17). As soon as the ices are brought up to this layer, the molecules will be liberated from the ices. On the other hand, the bottom panels show that re-freezing (i.e. due to downward vertical mixing, \citealt{KamaEA16}) occurs on very long timescales ($>$10$^{3}$ years) in the emitting layers, but occurs, as expected, very fast ($<$100 years) in the midplane of the dust trap. The vertical mixing timescale, based on turbulent diffusion, is dependent on the chosen value of $\alpha$. In Figure \ref{fig:Model-Diffusion}, we show the diffusion coefficients and corresponding timescales for values of $\alpha$=10$^{-4}$, 10$^{-3}$, and 10$^{-2}$. At the location of the dust trap, these values yield mixing timescales of, respectively, $t_\textnormal{mixing}\sim$(5-8)$\times$10$^5$ years, $\sim$(5-8)$\times$10$^4$ years, and $\sim$(5-8)$\times$10$^3$ years. In addition, we find diffusion coefficients (see also the top panels in Figure \ref{fig:Model-Diffusion}) in the emitting layer of $D_\textnormal{turbulence}\gtrsim$10$^{15}$ cm$^{2}$~s$^{-1}$ for the different values for $\alpha$, which is consistent with the typical value of $D_\textnormal{turbulence}\sim$10$^{17}$ cm$^{2}$~s$^{-1}$ for disks at distances of 100 au \citep{SemenovEA06}. \\
\indent Finally, Figure \ref{fig:Model-PD} shows the photodissociation timescales (top panels) and the distances the molecules can travel, under the assumption of a Keplerian orbit, before they are photodissociated (bottom panels). We find that the timescales are similar for all three molecules. For vertical heights of 10 to 15 au ($z/r\sim$0.17-0.25), the layer from which \ce{CH_3OH} and \ce{H_2CO} are expected to emit, the timescales are less than a year. Closer to the midplane, however, the timescales become a year and longer. These photodissociation timescales are significantly shorter than the vertical mixing timescales. Given the expected icy origin of the molecular emission, it is, therefore, unlikely that turbulent diffusion is the correct explanation for vertical mixing, unless the turbulence in the dust trap is really high and larger values for the $\alpha$-parameter ($\alpha$>0.01) should have been adopted. Other mechanisms, resulting favourably in faster timescales, need to be explored to further investigate the importance of vertical mixing and its dominant cause. One potential explanation is the vertical shear instability (VSI). The VSI is thought to more strongly mix the grains vertically than radially and should even be able to mix the grains into the warm molecular layers before molecules photodissociate \citep{FlockEA17,FlockEA20}. \\
\indent As the photodissociation timescales are very short in the emitting layers ($\leq$1 year), the molecules are not able to travel large distances before they are destroyed. However, for \ce{SO_2}, which lies deeper in the disk, there is a small layer above the midplane for the radial outer half of the dust trap, where the photodissociation timescales are long enough ($>$1 year) where the molecules can travel distances of $>$10 au to half an orbit (red line in the bottom panels of Figure \ref{fig:Model-PD}). This small layer of the dust trap agrees with our proposed emitting height based on the temperatures and the radial location as derived from the images. The larger distances the \ce{SO_2} molecules can travel in this layer may explain the azimuthal extent of the observed \ce{SO_2} transitions. For the higher proposed emitting layers of \ce{CH_3OH} and \ce{H_2CO}, the photodissociation timescales are too fast for the molecules to travel a significant distance. This can explain the azimuthally compact extent of the \ce{CH_3OH} emission, but not that of \ce{H_2CO}. \\
\indent Instead, we propose that the azimuthal extent of \ce{H_2CO} can be explained due to a significant contribution of gas-phase formation, which is efficient for \ce{H_2CO} but not for \ce{CH_3OH}. Gas-phase formation of \ce{H_2CO} is efficient in regions with efficient (ice) desorption and photodissociation \citep{AikawaEA02,LoomisEA15,CarneyEA17}, where atomic oxygen and radicals, such as \ce{CH_3}, \ce{CH_2}, and \ce{OH}, are present. The formation occurs through the reactions of \ce{CH_3}+\ce{O}$\rightarrow$\ce{H_2CO} + \ce{H} and \ce{CH_2}+\ce{OH}$\rightarrow$\ce{H_2CO}+\ce{H} \citep{FP02,AtkinsonEA06}. Interestingly, both \ce{CH_3} and \ce{OH} are photodissociation products of \ce{CH_3OH} \citep{KayanumaEA19}, therefore, the gas-phase formation of \ce{H_2CO} can be sustained by the photodissociation of, for example, \ce{CH_3OH}, \ce{H_2O} (which produces both \ce{OH} and \ce{O}), and \ce{CH_4} (which produces both \ce{CH_3} and \ce{CH_2}). In relation to our derived temperatures, we infer higher temperatures ($T>$300 K) for the centre of the dust trap and lower temperatures for the leading and trailing sides. This temperature difference may be explained by the ice-sublimated reservoir having a larger contribution in the centre, whereas the gas-phase formation reservoir may dominate more towards the sides of the dust trap. (Thermo-)chemical modelling efforts are required to further explore this scenario.

\section{Conclusions and summary} \label{sec:Conclusions}
With this work, we investigated the gas temperature of the asymmetric IRS~48 dust trap using the, respectively, 13, 22, and 7 observed transitions of \ce{SO_2}, \ce{CH_3OH}, and \ce{H_2CO} through pixel-by-pixel analyses. In addition, we provided new insights into the potential emitting locations of these species and determined their major timescales: vertical mixing, desorption, freeze-out, and photodissociation. Our conclusions are summarised as follows:
\begin{itemize}
    \item Our rotational diagram analysis resulted in temperatures of $T$=54.8$\pm$1.4 K and $T$=125.5$^{+3.7}_{-3.5}$ K at the peak positions of the \ce{SO_2} $J$=6$_{4,2}$-6$_{3,3}$ and \ce{CH_3OH} $J$=13$_{-1}$-13$_0$ transitions, respectively. In addition, our ratio maps for \ce{H_2CO} yield temperatures in the range of $T_\textnormal{rot}$=150 K to $T_\textnormal{rot}$=350 K.
    \item The well constrained rotational diagram of \ce{SO_2} suggests that the emission is thermalised, which occurs at gas densities of $n_\textnormal{gas}>$10$^8$ cm$^{-3}$. These densities and temperatures can only be found in the deepest layers of the disk, near the midplane (heights of $<$5 au or $z/r<$0.1), suggesting that the \ce{SO_2} emission originates from this layer. In contrast, the rotational diagram of \ce{CH_3OH} is characterised by a large scatter, which suggests sub-thermal excitation. Using non-LTE RADEX calculations we have shown that the gas densities for the derived temperatures at the locations of the dust trap are too low for the \ce{CH_3OH} and \ce{H_2CO} transitions to be thermalised. Therefore, our derived rotational temperatures underestimate the kinetic temperature. 
    \item The resulting temperature map of \ce{SO_2} hints at a potential radial gradient, where the temperature decreases with increasing radial distance. The map of \ce{CH_3OH} is, on the other hand, nearly uniform. We do not infer any hint of any possible vortex influencing the temperature structure of the dust trap. Given the rather low uncertainties of the derived temperatures, this influence, if present, is very small, on the order of a few K. The maps for \ce{H_2CO} show that the temperatures peak in the centre of the dust trap, while it decreases towards the leading and trailing edges.
    \item Additional analysis of the brightness temperature of the optically thick \ce{^{13}CO} $J$=3-2 and $J$=6-5 transitions, under the assumption that the emission is fully resolved, yield temperatures of  $T_\textnormal{b,3-2}\sim$110 K and $T_\textnormal{b,6-5}\sim$135 K. Combining these temperatures with the inferred rotational temperatures of \ce{SO_2}, \ce{CH_3OH}, and \ce{H_2CO}, we propose that the \ce{CH_3OH} and \ce{H_2CO} emission must originate from the highest elevated layer, also accounting for the rotational temperatures underestimate the kinetic ones. The lower brightness temperature of the \ce{^{13}CO} suggests that this molecule originates from deeper in the disk, whereas \ce{SO_2} likely emits from deep inside the disk near the midplane.
    \item Our pixel-by-pixel analysis suggests that some of the transitions of \ce{H_2CO} may be at most moderately optically thick, depending on the used temperature. In contrast, the previously low derived isotopic ratio observed for \ce{H_2CO} (\ce{H_2 ^{12}CO}/\ce{H_2 ^{13}CO}$\sim$4) suggests its the emission must be highly optically thick. For this, it must originate from an emitting area parametrised by a beam radius of $\sim$8 au. Such a small beam radius suggests that the \ce{H_2CO} emission must originate from a small radial ``sliver'' at the inner edge of the dust trap. As the current observations do not radially resolve the emission, higher resolution observations are required to fully constrain the optical depth of \ce{H_2CO}.
    \item Finally, we have determined the turbulent vertical mixing, desorption, freeze-out, and photodissociation timescales at the location of the dust trap. The photodissociation timescales are too fast compared with the turbulent mixing timescales, therefore turbulent diffusion cannot explain the observed molecular emission. Other vertical mixing mechanisms should be explored to investigate its influence on the observable chemistry. 
    \item The photodissociation timescale of \ce{SO_2} and the corresponding azimuthal distance the molecules can travel before they are destroyed, under the assumption of a Keplerian orbit, can explain its observed azimuthal extent (up to approximately a quarter of an orbit) of the molecular emission. The compact azimuthal extent of the \ce{CH_3OH} is, on the other hand, consistent with the short photodissociation timescales in the expected emitting layer, higher up in the disk.
    \item Photodissociation occurs too fast to explain the azimuthal extent of \ce{H_2CO}. Instead, we propose that the extent can be explained by an additional reservoir that follows from gas-phase reactions, which may be sustained by the photodissociation of \ce{CH_3OH} and \ce{H_2O}. The large inferred temperature range, $T_\textnormal{rot}\sim$150-350 K, can also be explained by these two reservoirs: the gas-phase reservoir dominates in the cold outer regions of the dust trap, whereas the reservoir resulting from the sublimation of ices mainly contributes to the hot central region.
\end{itemize}
Our findings highlight the power of line surveys and the need for multiple transitions of the same molecule for the characterisation of the emission in disks and an empirical determination of the gas temperature across and within the disk. Our analysis of the IRS~48 disk shows that the properties of the molecular emission are more complex than originally thought, with both the \ce{CH_3OH} and \ce{H_2CO} being sub-thermally excited and that the low isotopologue ratio for \ce{H_2CO} can only be explained if optically thick emission originates from a small radial ``sliver''. Higher spatial and spectral resolution observations are required to further investigate the emitting regions of each molecular species and, therefore, gain more insights into the radial width over which sublimation is important in the dust traps of transition disks.

\begin{acknowledgements}
    The authors thank Haifeng Yang for sharing the high-resolution continuum observations of IRS~48. The authors acknowledge assistance from Allegro, the European ALMA Regional Center node in the Netherlands. \\
    \indent This paper makes use of the following ALMA data: 2013.1.00100.S, 2017.1.00834.S, 2019.1.01059.S, 2021.1.00738.S. ALMA is a partnership of ESO (representing its member states), NSF (USA) and NINS (Japan), together with NRC (Canada), NSTC and ASIAA (Taiwan), and KASI (Republic of Korea), in cooperation with the Republic of Chile. The Joint ALMA Observatory is operated by ESO, AUI/NRAO and NAOJ. \\
    \indent M.T. and E.F.v.D. acknowledge support from the ERC grant 101019751 MOLDISK. A.S.B is supported by a Clay Postdoctoral Fellowship from the Smithsonian Astrophysical Observatory. M.L. is funded by the European Union (ERC, UNVEIL, 101076613). Views and opinions expressed are however those of the author(s) only and do not necessarily reflect those of the European Union or the European Research Council. Neither the European Union nor the granting authority can be held responsible for them. E.F.v.D. acknowledges support from the Danish National Research Foundation through the Center of Excellence ``InterCat'' (DNRF150). L.E. acknowledges financial support from the Science and Technology Facilities Council (grant number ST/X001016/1). L.K. is funded by UKRI guaranteed funding for a Horizon Europe ERC consolidator grant (EP/Y024710/1). Support for C.J.L. was provided by NASA through the NASA Hubble Fellowship grant No. HST-HF2-51535.001-A awarded by the Space Telescope Science Institute, which is operated by the Association of Universities for Research in Astronomy, Inc., for NASA, under contract NAS5-26555. S.N. is grateful for support from Grants-in-Aid for JSPS (Japan Society for the Promotion of Science) Fellows Grant Number JP23KJ0329, and MEXT/JSPS Grants-in-Aid for Scientific Research (KAKENHI) Grant Numbers JP20H05845, JP20H05847, JP23K13155, and JP24K00674. C.W.~acknowledges financial support from the Science and Technology Facilities Council and UK Research and Innovation (grant numbers ST/X001016/1 and MR/T040726/1). \\
    \indent The project leading to this publication has received support from ORP, that is funded by the European Union's Horizon 2020 research and innovation programme under grant agreement No 101004719 [ORP]. \\
    \indent This work also has made use of the following software packags that have not been mentioned in the main text: NumPy, SciPy, Astropy, Matplotlib, pandas, IPython, Jupyter \citep{Numpy,Scipy,AstropyI,AstropyII,AstropyIII,Matplotlib,pandas,IPython,Jupyter}.
\end{acknowledgements}

\bibliographystyle{aa}
\bibliography{Bibliography}


\onecolumn
\begin{appendix}
\section{\ce{SO_2}, \ce{CH_3OH}, and \ce{H_2CO} transitions}
\begin{table*}[ht!]
    \centering
    \caption{The detected transitions of \ce{SO_2} (top part), \ce{CH_3OH} (middle part), and \ce{H_2CO} (bottom part).}
    \begin{tabular}{c c c c c c c c c}
        \hline
        \hline
        Molecule & Transition$^{(a)}$ & Frequency & E$_\textnormal{low}$ & E$_\textnormal{up}$ & A$_\textnormal{ul}$ & g$_\textnormal{low}$ & g$_\textnormal{up}$ & Int. flux \\
         & & [GHz] & [K] & [K] & [s$^{-1}$] & & & [mJy~beam$^{-1}$ km~s$^{-1}$] \\
        \hline
        \ce{SO_2}$^{(b)}$ & 18$_{4,14}$-18$_{3,15}$ & 338.3059931 & 180.6 & 196.8 & 3.3$\times$10$^{-4}$ & 37 & 37 & 16.1$\pm$2.7 \\
                  & 13$_{2,12}$-12$_{1,11}$ & 345.3385377 & 76.4 & 93.0 & 2.4$\times$10$^{-4}$ & 25 & 27 & 58.9$\pm$2.0 \\
                  & 5$_{3,3}$-4$_{2,2}$ & 351.2572233 & 19.0 & 35.9 & 3.4$\times$10$^{-4}$ & 9 & 11 & 70.5$\pm$2.6 \\
                  & 14$_{4,10}$-14$_{3,11}$ & 351.8738732 & 119.0 & 135.9 & 3.4$\times$10$^{-4}$ & 29 & 29 & 34.9$\pm$2.6 \\
                  & 12$_{4,8}$-12$_{3,9}$ & 355.0455169 & 94.0 & 111.0 & 3.4$\times$10$^{-4}$ & 25 & 25 & 44.7$\pm$2.2 \\
                  & 10$_{4,6}$-10$_{3,7}$ & 356.7551899 & 72.7 & 89.8 & 3.3$\times$10$^{-4}$ & 21 & 21 & 49.1$\pm$2.7 \\
                  & 13$_{4,10}$-13$_{3,11}$ & 357.1653904 & 105.8 & 123.0 & 3.5$\times$10$^{-4}$ & 27 & 27 & 39.6$\pm$2.5 \\
                  & 15$_{4,12}$-15$_{3,13}$ & 357.2411932 & 132.5 & 149.7 & 3.6$\times$10$^{-4}$ & 31 & 31 & 35.0$\pm$2.9 \\
                  & 11$_{4,8}$-11$_{3,9}$ & 357.3875795 & 82.8 & 100.0 & 3.4$\times$10$^{-4}$ & 23 & 23 & 48.8$\pm$2.3 \\
                  & 8$_{4,4}$-8$_{3,5}$ & 357.5814486 & 55.2 & 72.4 & 3.1$\times$10$^{-4}$ & 17 & 17 & 47.6$\pm$2.5 \\
                  & 9$_{4,6}$-9$_{3,7}$ & 357.6718206 & 63.5 & 80.6 & 3.2$\times$10$^{-4}$ & 19 & 19 & 46.9$\pm$2.5 \\
                  & 7$_{4,4}$-7$_{3,5}$ & 357.8924422 & 47.8 & 65.0 & 2.9$\times$10$^{-4}$ & 15 & 15 & 50.3$\pm$2.3 \\
                  & 6$_{4,2}$-6$_{3,3}$ & 357.9258478 & 41.4 & 58.6 & 2.6$\times$10$^{-4}$ & 13 & 13 & 55.3$\pm$2.7 \\
        \hline
        \ce{CH_3OH}$^{(b)}$ & 3$_{3}$-4$_{2}$ (E) & 337.135853 & 45.5 & 61.6 & 1.6$\times$10$^{-5}$ & 9 & 7 & 20.0$\pm$3.4 \\
                    & 7$_{0}$-6$_{0}$ (E) & 338.124488 & 61.9 & 78.1 & 1.7$\times$10$^{-4}$ & 13 & 15 & 67.5$\pm$4.1 \\
                    & 7$_{-1}$-6$_{-1}$ (E) & 338.344588 & 54.3 & 70.6 & 1.7$\times$10$^{-4}$ & 13 & 15 & 66.7$\pm$3.8 \\
                    & 7$_{-6}$-6$_{-6}$ (E) & 338.430975 & 237.7 & 253.9 & 4.5$\times$10$^{-5}$ & 13 & 15 & 9.7$\pm$2.5 \\
                    & 7$_{-6}$-6$_{-6}$ (A) & 338.442367 & 242.5 & 258.7 & 4.5$\times$10$^{-5}$ & 13 & 15 & 14.3$\pm$3.2 \\
                    & 7$_{6}$-6$_{6}$ (A) & 338.442367 & 242.5 & 258.7 & 4.5$\times$10$^{-5}$ & 13 & 15 & 14.3$\pm$3.2 \\
                    & 7$_{-5}$-6$_{-5}$ (E) & 338.456536 & 172.8 & 189.0 & 8.3$\times$10$^{-5}$ & 13 & 15 & 22.6$\pm$3.2 \\
                    & 7$_{5}$-6$_{5}$ (E) & 338.475226 & 184.8 & 201.1 & 8.4$\times$10$^{-5}$ & 13 & 15 & 33.4$\pm$3.4 \\
                    & 7$_{5}$-6$_{5}$ (A) & 338.486322 & 186.6 & 202.9 & 8.4$\times$10$^{-5}$ & 13 & 15 & 18.6$\pm$3.1 \\
                    & 7$_{-5}$-6$_{-5}$ (A) & 338.486322 & 186.6 & 202.9 & 8.4$\times$10$^{-5}$ & 13 & 15 & 18.6$\pm$3.1 \\
                    & 7$_{-4}$-6$_{-4}$ (E) & 338.504065 & 136.6 & 152.9 & 1.1$\times$10$^{-4}$ & 13 & 15 & 42.1$\pm$3.6 \\
                    & 7$_{4}$-6$_{4}$ (E) & 338.530257 & 144.7 & 161.0 & 1.2$\times$10$^{-4}$ & 13 & 15 & 43.8$\pm$3.8 \\
                    & 7$_{-3}$-6$_{-3}$ (E) & 338.559963 & 111.5 & 127.7 & 1.4$\times$10$^{-4}$ & 13 & 15 & 48.4$\pm$3.6 \\
                    & 7$_{3}$-6$_{3}$ (E) & 338.583216 & 96.5 & 112.7 & 1.4$\times$10$^{-4}$ & 13 & 15 & 56.9$\pm$3.8 \\
                    & 7$_{2}$-6$_{2}$ (A) & 338.639802 & 86.5 & 102.7 & 1.6$\times$10$^{-4}$ & 13 & 15 & 57.3$\pm$3.4 \\
                    & 2$_{2}$-3$_{1}$ (A) & 340.141143 & 28.3 & 44.7 & 2.8$\times$10$^{-5}$ & 7 & 5 & 13.2$\pm$2.8 \\
                    & 13$_{-1}$-13$_{0}$ (A) & 342.729796 & 211.0 & 227.5 & 4.2$\times$10$^{-4}$ & 27 & 27 & 64.0$\pm$3.0 \\
                    & 14$_{-1}$-14$_{0}$ (A) & 349.106997 & 243.4 & 260.2 & 2.2$\times$10$^{-4}$ & 29 & 29 & 58.7$\pm$3.0 \\
                    & 1$_{1}$-0$_{0}$ (A) & 350.905100 & 0.0 & 16.8 & 3.3$\times$10$^{-4}$ & 1 & 3 & 51.2$\pm$3.5 \\
                    & 9$_{5}$-10$_{4}$ (E) & 351.236479 & 223.6 & 240.5 & 3.6$\times$10$^{-5}$ & 21 & 19 & 25.3$\pm$2.8 \\
                    & 13$_{0}$-12$_{1}$ (A) & 355.602945 & 194.0 & 211.0 & 1.3$\times$10$^{-4}$ & 25 & 27 & 61.9$\pm$3.3 \\
                    & 15$_{-1}$-15$_{0}$ (A) & 356.007235 & 278.2 & 295.3 & 2.3$\times$10$^{-4}$ & 31 & 31 & 49.6$\pm$3.0 \\
        \hline
        \ce{H_2CO}$^{(b)}$ & 5$_{1,5}$-4$_{1,4}$ (o) & 351.768645 & 45.6 & 62.5 & 12.0$\times$10$^{-4}$ & 27 & 33 & 265.7$\pm$5.8 \\
                   & 5$_{0,5}$-4$_{0,4}$ (p) & 362.736048 & 34.9 & 52.3 & 13.7$\times$10$^{-4}$ & 9 & 11 & 188.9$\pm$7.1 \\
                   & 5$_{2,4}$-4$_{2,3}$ (p) & 363.945894 & 82.1 & 99.5 & 11.6$\times$10$^{-4}$ & 9 & 11 & 167.4$\pm$8.1 \\
                   & 5$_{4,2}$-4$_{4,1}$ (p) & 364.103249 & 223.3 & 240.7 & 5.0$\times$10$^{-4}$ & 9 & 11 & 53.6$\pm$8.1 \\
                   & 5$_{4,1}$-4$_{4,0}$ (p) & 364.103249 & 223.3 & 240.7 & 5.0$\times$10$^{-4}$ & 9 & 11 & 53.6$\pm$8.1 \\
                   & 5$_{3,3}$-4$_{3,2}$ (o) & 364.275141 & 140.9 & 158.4 & 8.9$\times$10$^{-4}$ & 27 & 33 & 169.9$\pm$7.6 \\
                   & 5$_{3,2}$-4$_{3,1}$ (o) & 364.288884 & 140.9 & 158.4 & 8.9$\times$10$^{-4}$ & 27 & 33 & 175.8$\pm$8.5 \\
        \hline
    \end{tabular}
    \tablefoot{$^{(a)}$: The listed transitions consist of the quantum configuration as given in the LAMDA files: $J_{K_a,K_c}$ for \ce{SO_2}, $J_K$ for \ce{CH_3OH}, and $J_{K_p,K_o}$ for \ce{H_2CO}. \\
    $^{(b)}$: The beam sizes for the \ce{SO_2} and \ce{CH_3OH} transitions are on the order of 0.33"$\times$0.26", whereas those for \ce{H_2CO} are on the order of 0.48"$\times$0.38".}
    \label{tab:PeakDetections}
\end{table*}

\begin{table*}[ht!]
    \centering
    \caption{All removed transitions due to blending and/or noise.}
    \begin{tabular}{c c c c c c}
        \hline
        \hline
        Molecule & Transition$^{(a)}$ & Frequency & E$_\textnormal{up}$ & $A_\textnormal{ul}$ & Reason \\
                 &                    & [GHz]     & [K]                 & [s$^{-1}$]          &        \\
        \hline
        \multirow{7}{1cm}{\ce{SO_2}} & 20$_{1,19}$-19$_{2,18}$ & 338.6118103 & 198.9 & 2.9$\times$10$^{-4}$ & Blend \\
                                     & 28$_{2,16}$-28$_{1,27}$ & 340.3164059 & 391.8 & 2.6$\times$10$^{-4}$ & Noise \\
                                     & 29$_{3,27}$-30$_{0,30}$ & 342.634969 & 421.4 & 2.5$\times$10$^{-6}$ & Noise \\
                                     & 5$_{5,1}$-6$_{4,2}$ & 345.1489708 & 75.1 & 9.8$\times$10$^{-6}$ & Noise \\
                                     & 10$_{6,4}$-11$_{5,7}$ & 350.862756 & 138.8 & 4.4$\times$10$^{-5}$ & Noise \\
                                     & 17$_{4,14}$-18$_{1,17}$ & 355.1864962 & 180.1 & 2.6$\times$10$^{-6}$ & Noise \\
                                     & 15$_{7,9}$-16$_{6,10}$ & 356.0406442 & 230.4 & 6.4$\times$10$^{-5}$ & Noise \\
        \hline
        \multirow{11}{1cm}{\ce{CH_3OH}} & 7$_{6}$-6$_{6}$ (E) & 338.40461 & 243.8 & 4.5$\times$10$^{-5}$ & Blend \\
                                        & 7$_{0}$-6$_{0}$ (A) & 338.408698 & 65.0 & 1.7$\times$10$^{-4}$ & Blend \\
                                        & 7$_{1}$-6$_{1}$ (E) & 338.614936 & 86.1 & 1.7$\times$10$^{-4}$ & Blend \\
                                        & 7$_{2}$-6$_{2}$ (E) & 338.721693 & 87.3 & 1.6$\times$10$^{-4}$ & Blend \\
                                        & 7$_{-2}$-6$_{-2}$ (E) & 338.722898 & 90.9 & 1.6$\times$10$^{-4}$ & Blend \\
                                        & 7$_{-4}$-6$_{-4}$ (A) & 338.512632 & 145.3 & 1.1$\times$10$^{-4}$ & Blend \\
                                        & 7$_{4}$-6$_{4}$ (A) & 338.512644 & 145.3 & 1.1$\times$10$^{-4}$ & Blend \\
                                        & 7$_{-2}$-6$_{-2}$ (A) & 338.512853 & 102.7 & 1.6$\times$10$^{-4}$ & Blend \\
                                        & 7$_{3}$-6$_{3}$ (A) & 338.540826 & 114.8 & 1.4$\times$10$^{-4}$ & Blend \\
                                        & 7$_{-3}$-6$_{-3}$ (A) & 338.543152 & 114.8 & 1.4$\times$10$^{-4}$ & Blend \\
                                        & 4$_{0}$-3$_{-1}$ (E) & 350.687662 & 36.3 & 8.7$\times$10$^{-5}$ & Blend \\
        \hline
    \end{tabular}
    \tablefoot{$^{(a)}$: The listed transitions consist of the quantum configuration as given in the LAMDA files: $J_{K_a,K_c}$ for \ce{SO_2} and $J_K$ for \ce{CH_3OH}.}
    \label{tab:RemTrans}
\end{table*}

\clearpage
\onecolumn
\section{Line profiles}
\begin{figure*}[ht!]
    \centering
    \includegraphics[width=\textwidth]{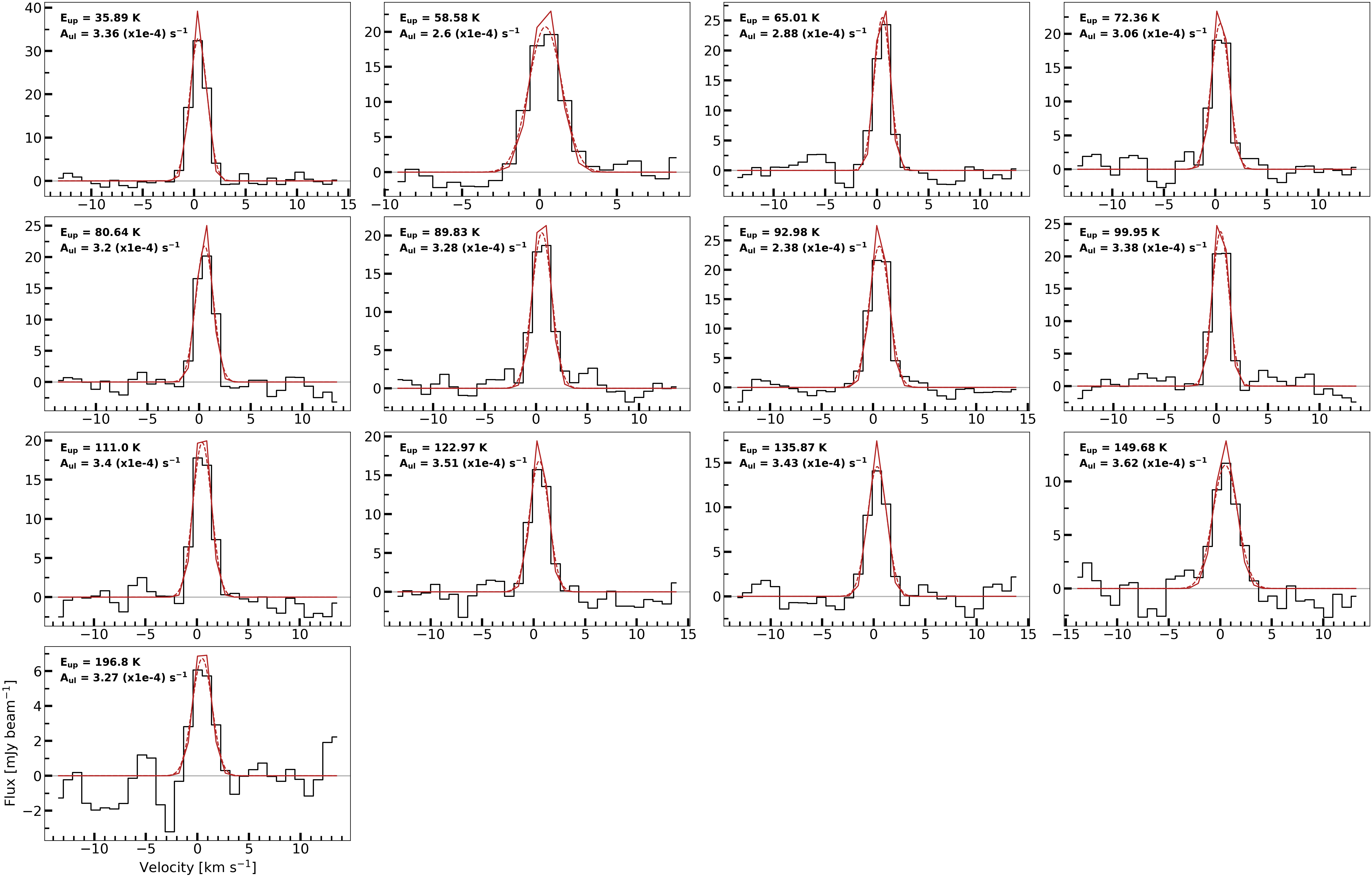}
    \caption{The line profiles of the detected \ce{SO_2} transitions at the peak flux location of the 6$_{4,2}$-6$_{3,3}$ transition. The red line displays the fitted Gaussian line profile at the resolution of the spectrum (0.9 km s$^{-1}$), whereas the red dashed line shows the Gaussian line profile at a higher resolution. The lines are ordered based on their upper level energies ($E_\textnormal{up}$).}
    \label{fig:SO2-Lines}
\end{figure*}

\clearpage
\begin{figure*}[ht!]
    \centering
    \includegraphics[width=\textwidth]{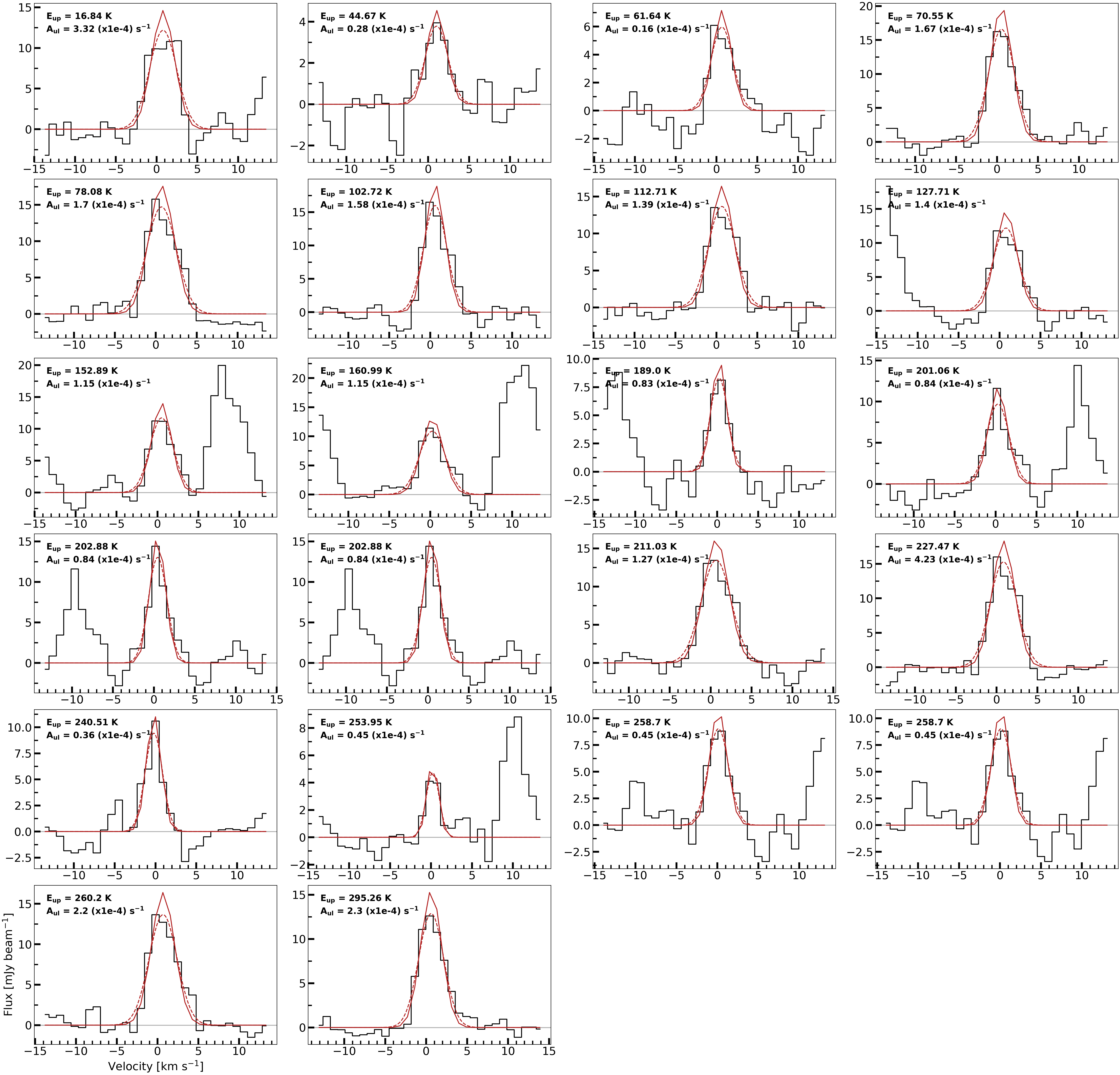}
    \caption{Similar as Figure \ref{fig:SO2-Lines}, but for the detected \ce{CH_3OH} transitions at the peak flux position of the 13$_{-1}$-13$_0$ transition.}
    \label{fig:CH3OH-Lines}
\end{figure*}

\clearpage
\begin{figure*}[ht!]
    \centering
    \includegraphics[width=\textwidth]{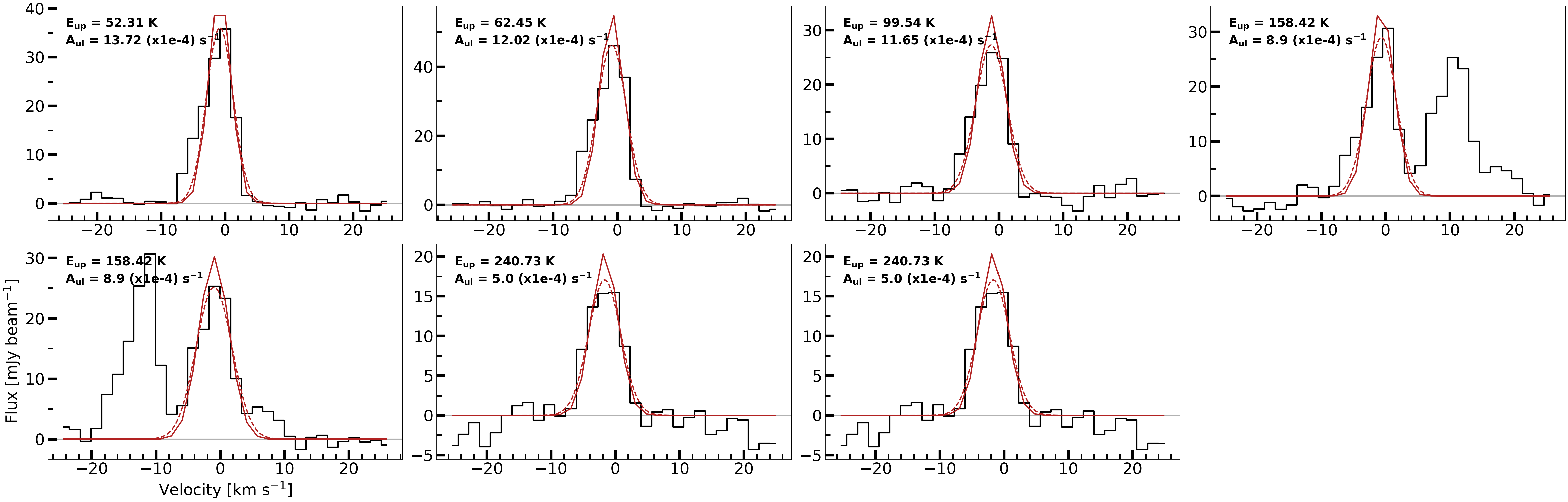}
    \caption{Similar as Figure \ref{fig:SO2-Lines}, but for the detected \ce{H_2CO} transitions at the peak flux position of the 5$_{0,5}$-4$_{0,4}$ transition.}
    \label{fig:H2CO-Lines}
\end{figure*}

\section{\ce{H_2CO} line ratios} \label{sec:H2CO-LRs}
\textit{5$_\mathit{3,3}$-4$_\mathit{3,2}$/5$_\mathit{1,5}$-4$_\mathit{1,4}$}: for the ortho-transition line pair, we see a clear spot (bright yellow) in the ratio map, where the ratios exceed those from the RADEX calculations. Therefore, the temperature derived for these pixels is limited by the upper range of 500 K. Ratios involving transitions with higher upper level energies are, therefore, needed to infer the temperature for these pixels. For the pixels for which the ratio does not exceed those from the calculations, we generally find temperatures, for both gas densities (and, therefore, across the vertical height of the disk), on the order of $\sim$150-300 K. These are similar to and slightly higher than the derived temperatures for the disk-integrated fluxes of $\sim$200 K inferred by \citet{vdMarelEA21} (red line in the left panels). The peak pixel of the 5$_{0,5}$-4$_{0,4}$ transition is contained within the region of pixels that exceed the ratios from the RADEX calculations and the temperature is, therefore, not well constrained. \\
\indent \textit{5$_\mathit{2,4}$-4$_\mathit{2,3}$/5$_\mathit{0,5}$-4$_\mathit{0,4}$}: for this first para-transition line pair we find that the inferred ratios from the observations exceed the ratios obtained from the RADEX calculations. This is likely the result of the low upper level energies of the transitions involved, 100 K and 52 K, respectively, which are not sensitive to the higher temperatures $\sim$150-350 K as probed by the ortho line pair. \\
\indent \textit{5$_\mathit{4,2}$-4$_\mathit{4,1}$/5$_\mathit{0,5}$-4$_\mathit{0,4}$}: for the second line pair, we find very similar results to the single ortho line pair. We find a spot in the centre of the maps that is limited by the temperature sensitivity of the lines. In addition, we find temperatures, for the other pixels, on the order of 150 K to 300 K for both gas densities. Those temperatures are, again, similar to those inferred for the disk-integrated fluxes by \citet{vdMarelEA21}. Once more, the peak position of the 5$_{0,5}$-4$_{0,4}$ transition falls within the range for which the temperature is not well constrained. \\
\indent \textit{5$_\mathit{4,2}$-4$_\mathit{4,1}$/5$_\mathit{2,4}$-4$_\mathit{2,3}$}: for this final line pair, which involves the ratio of the transitions with the highest upper level energies (241 K and 100 K, respectively), we find similar results as for the ratio (5$_\mathit{4,2}$-4$_\mathit{4,1}$/5$_\mathit{0,5}$-4$_\mathit{0,4}$) above, although we are able to provide better constraints on the temperature for a larger number of pixels. The temperature across the disk, especially when considering the gas density of $n_\textnormal{\ce{H_2}}$=10$^8$ cm$^{-3}$, is on the order of 150 K to 350 K. Again, this is similar to or slightly higher than the disk-integrated one.

\clearpage
\section{Observations-to-model ratios and uncertainty maps}
\begin{figure*}[ht!]
    \centering
    \includegraphics[width=\textwidth]{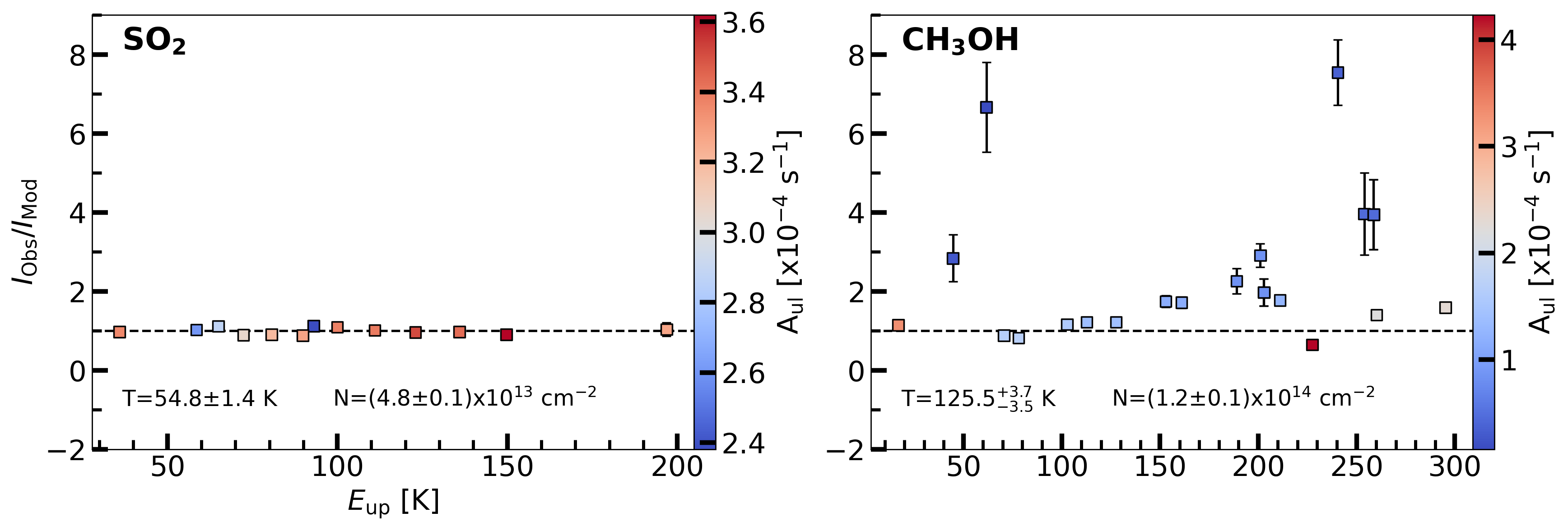}
    \caption{Intensity ratios between the observations and the model for \ce{SO_2} (left) and \ce{CH_3OH} (right). The horizontal bar in the ratio plots indicates a value of unity. The colour bar indicates the value for $A_\textnormal{ul}$ of the transitions.}
    \label{fig:RDAs-Ratios}
\end{figure*}
\begin{figure*}[ht!]
    \centering
    \includegraphics[width=\textwidth]{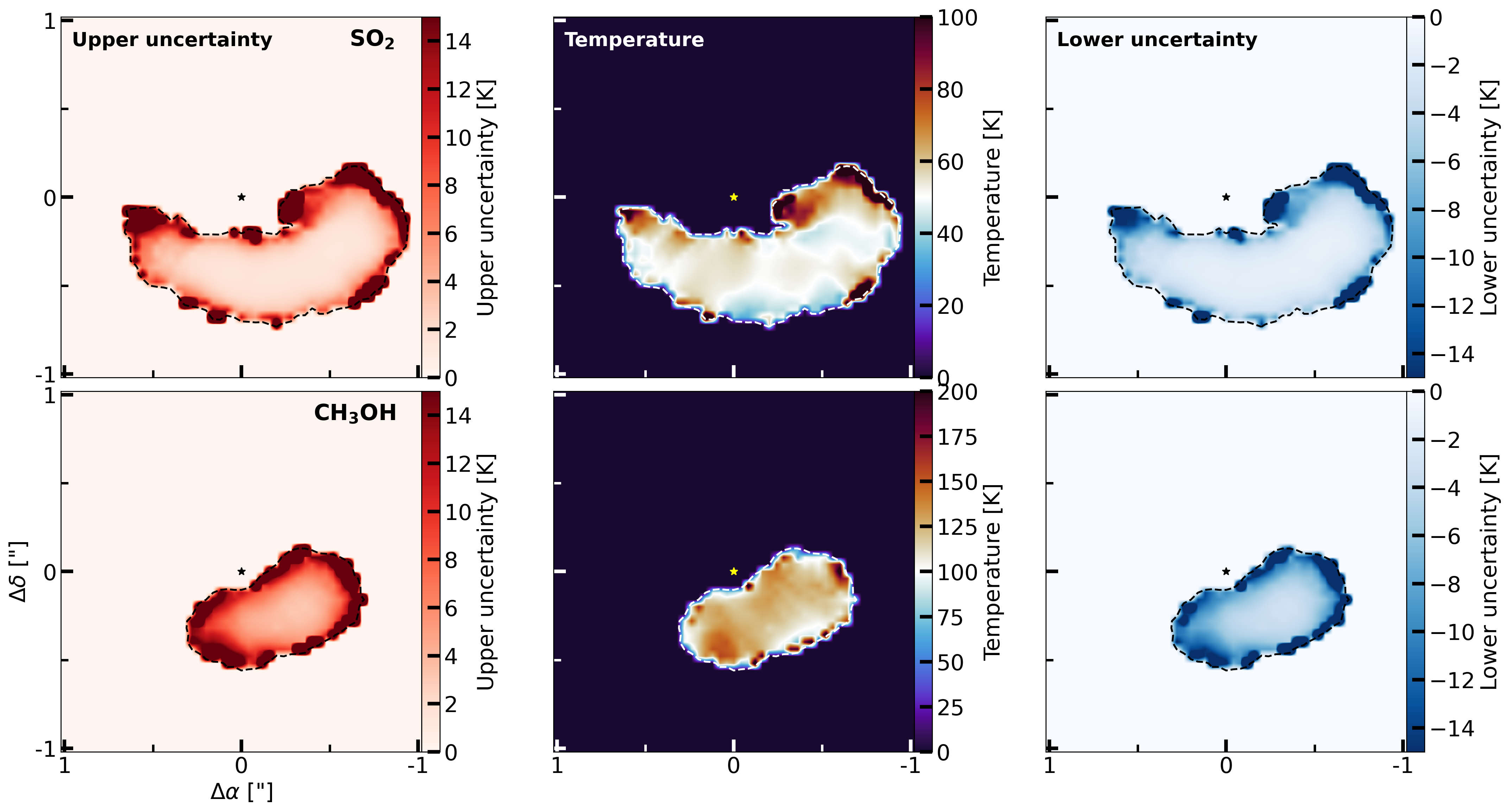}
    \caption{Uncertainty maps for the derived rotational temperatures. The left panels displays the upper uncertainties, while the right panels show the lower uncertainties. The derived temperature maps themselves are shown in the middle panels. The top row corresponds to \ce{SO_2}, whereas the bottom row displays those for \ce{CH_3OH}.}
    \label{fig:TUncertainty}
\end{figure*}
\begin{figure*}[ht!]
    \centering
    \includegraphics[width=\textwidth]{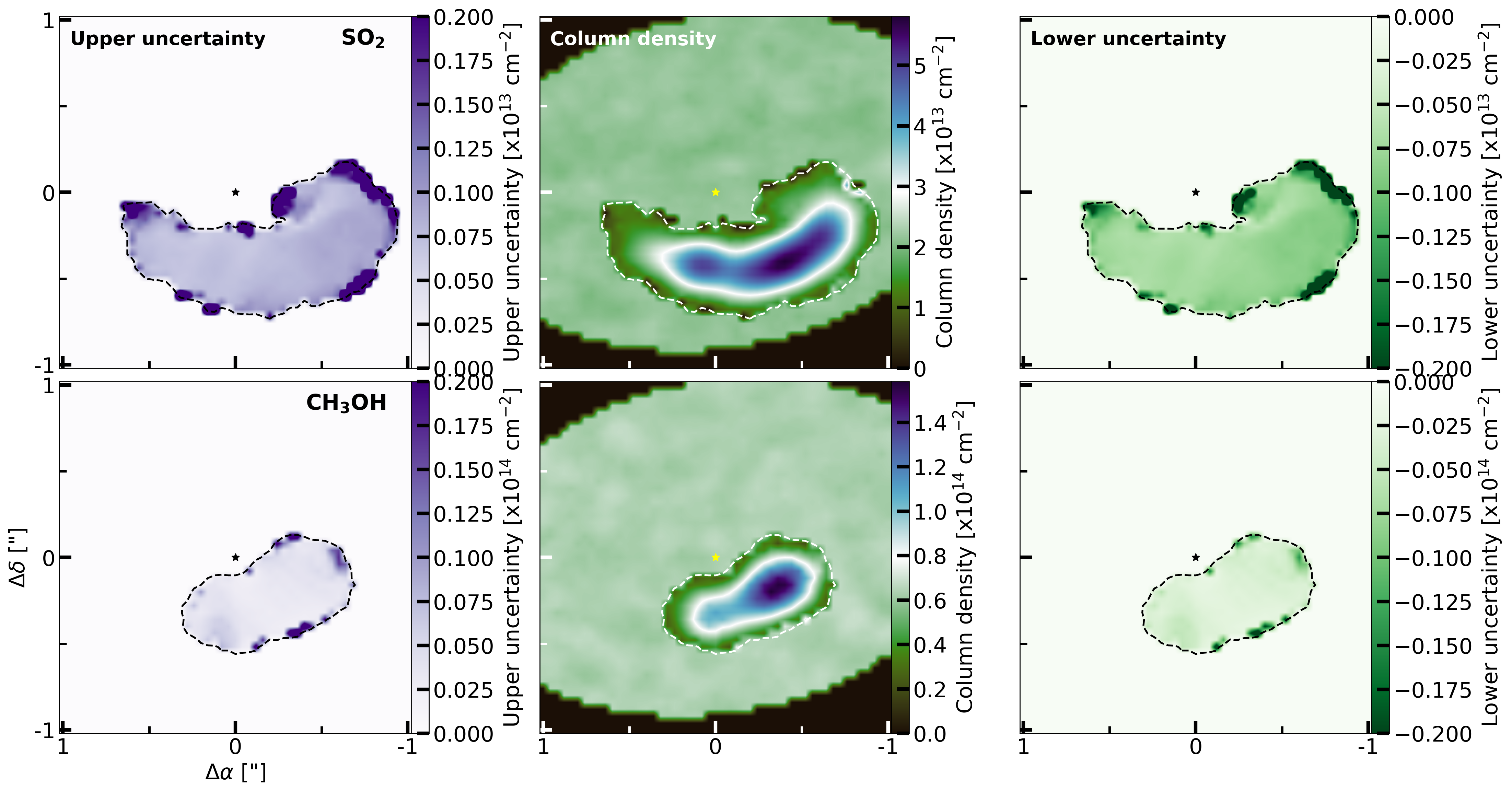}
    \caption{Similar as Figure \ref{fig:TUncertainty}, but for the column density maps.}
    \label{fig:NUncertainty}
\end{figure*}

\clearpage
\section{\ce{H_2CO} moment-0 maps}
\begin{figure*}[ht!]
    \centering
    \includegraphics[width=\textwidth]{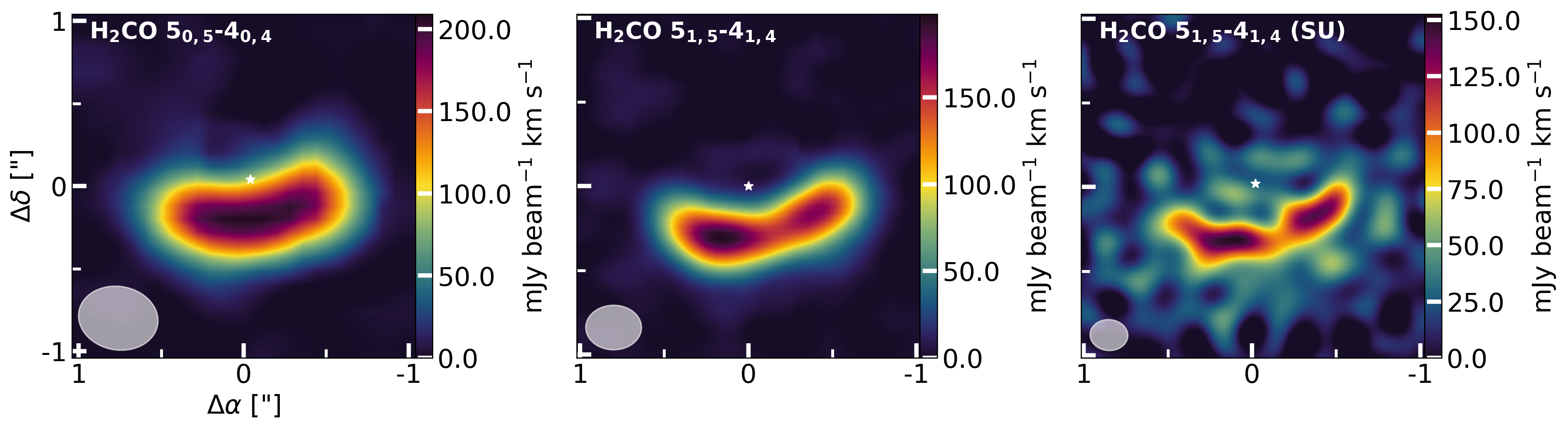}
    \caption{Moment-0 maps of the \ce{H_2CO} $J$=5$_{0,5}$-4$_{0,4}$ transition from the low resolution observations of \citet{vdMarelEA21} (left) and the $J$=5$_{1,5}$-4$_{1,4}$ transition from the higher resolution observations of \citet{BoothEA24}, imaged both with Briggs weighting (robust of 0.5, middle panel) and superuniform weighting (right). The white star in the centre represent the approximate location of the host star, while the resolving beam is shown in the lower left corner.}
    \label{fig:M0s-H2CO}
\end{figure*}

\section{Observed transitions per pixel}
\begin{figure*}[ht!]
    \centering
    \includegraphics[width=\textwidth]{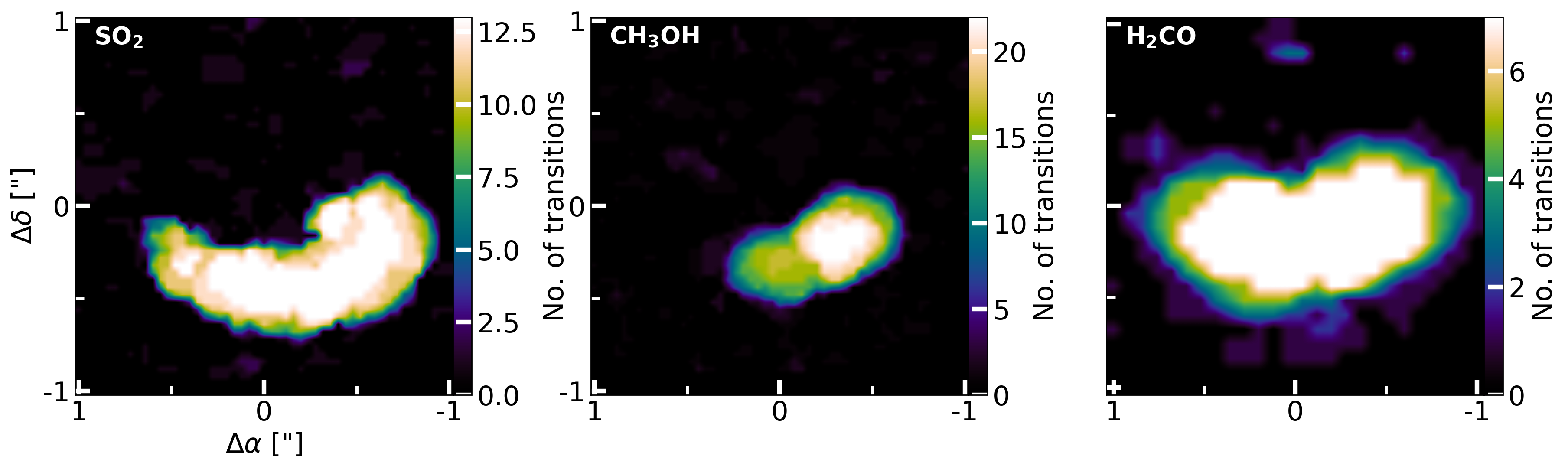}
    \caption{Number of detected transitions for \ce{SO_2} (left), \ce{CH_3OH} (middle), and \ce{H_2CO} (right) across the dust trap.}
    \label{fig:NoDetections}
\end{figure*}

\clearpage
\section{Channel maps}
\begin{figure*}[ht!]
    \centering
    \includegraphics[trim=0cm 0cm 0cm 1cm, clip=true, width=0.8\textwidth]{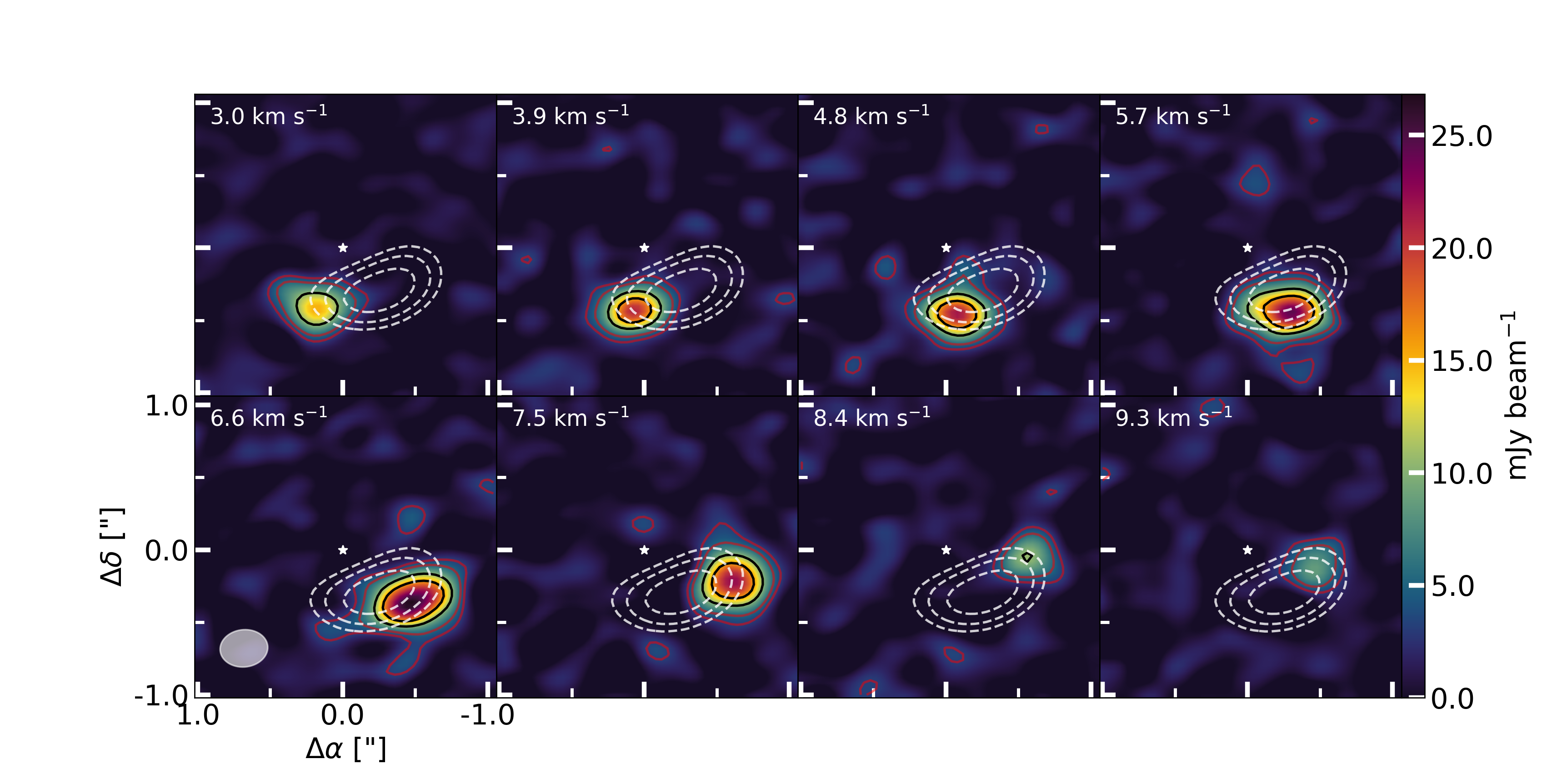}
    \includegraphics[trim=0cm 0cm 0cm 1cm, clip=true, width=0.8\textwidth]{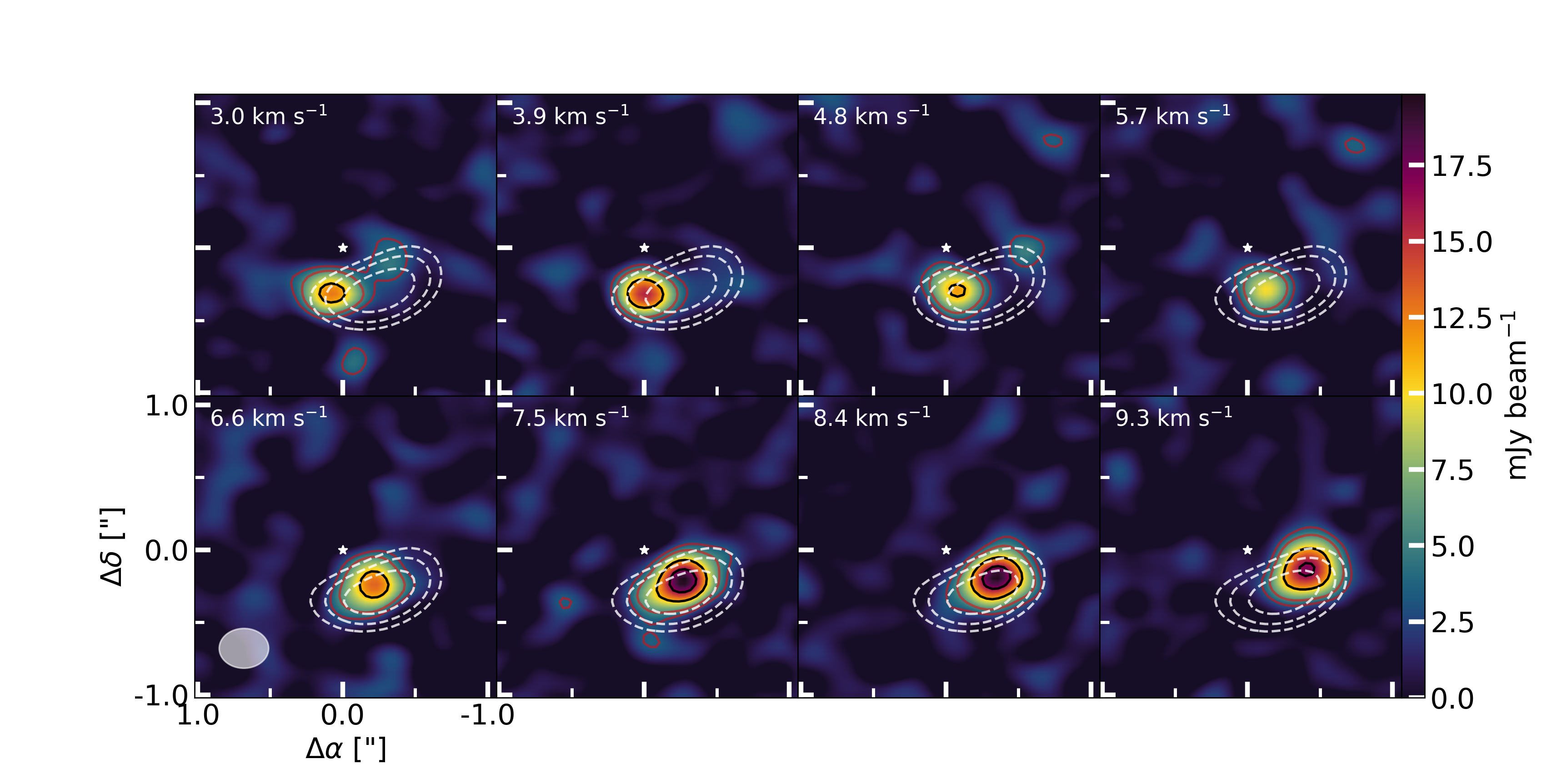}
    \includegraphics[trim=0cm 0cm 0cm 1cm, clip=true, width=0.8\textwidth]{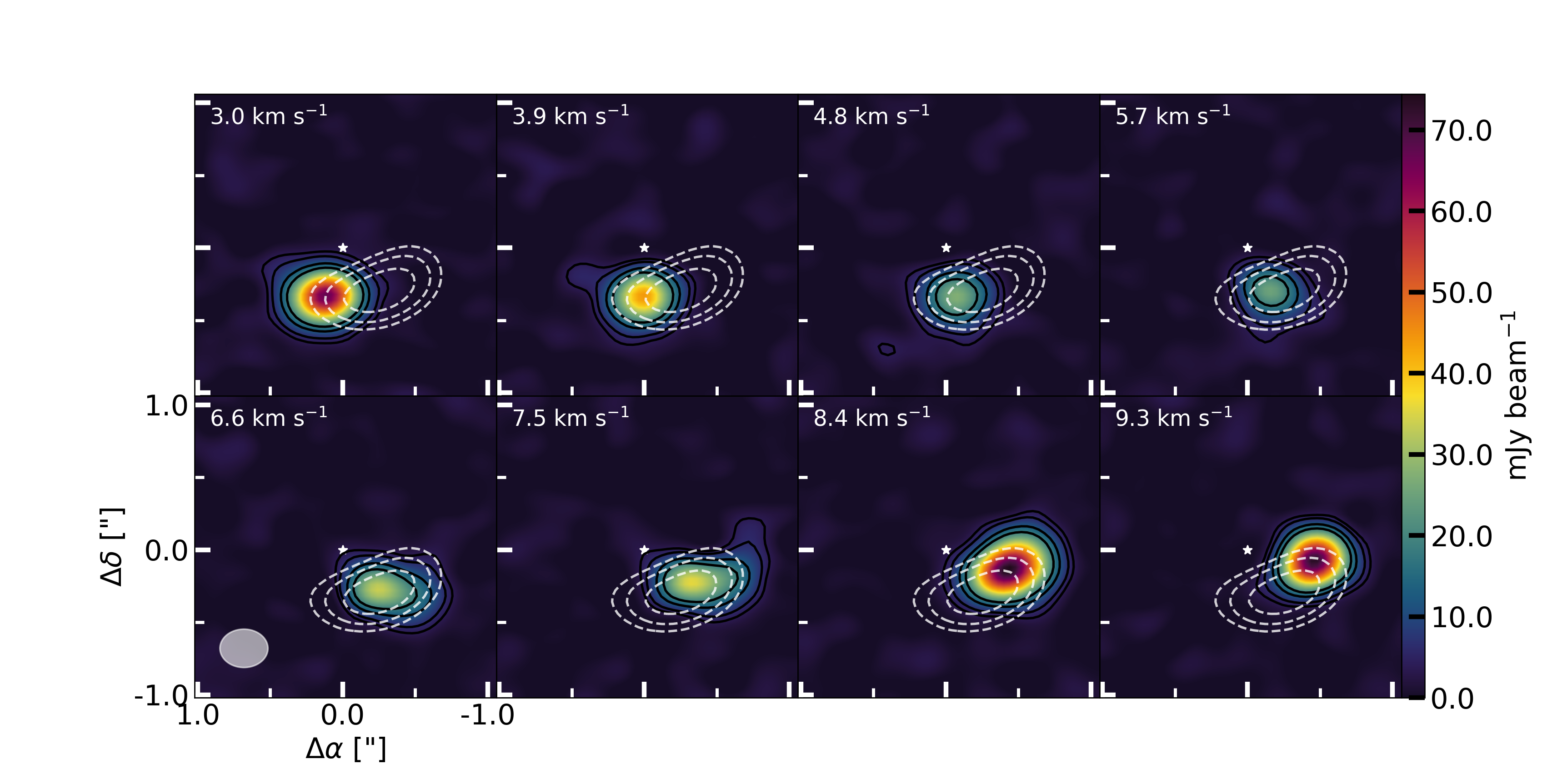}
    \caption{Channel maps of \ce{SO_2} (top row), \ce{CH_3OH} (middle row), and \ce{H_2CO} (bottom row) over the 3.0-9.3 km s$^{-1}$ range. The solid contours indicate, the 3$\times$, 5$\times$ (both in red), 10$\times$ and 15$\times$ (both in black) RMS levels of the emission, whereas the white dashed contours indicate the location of the continuum emission. The white star indicates the location of the host stars, whereas the grey ellipse denotes the beam.}
    \label{fig:ChannelMaps}
\end{figure*}

\clearpage
\section{Timescale figures}
\begin{figure*}[ht!]
    \centering
    \includegraphics[width=\textwidth]{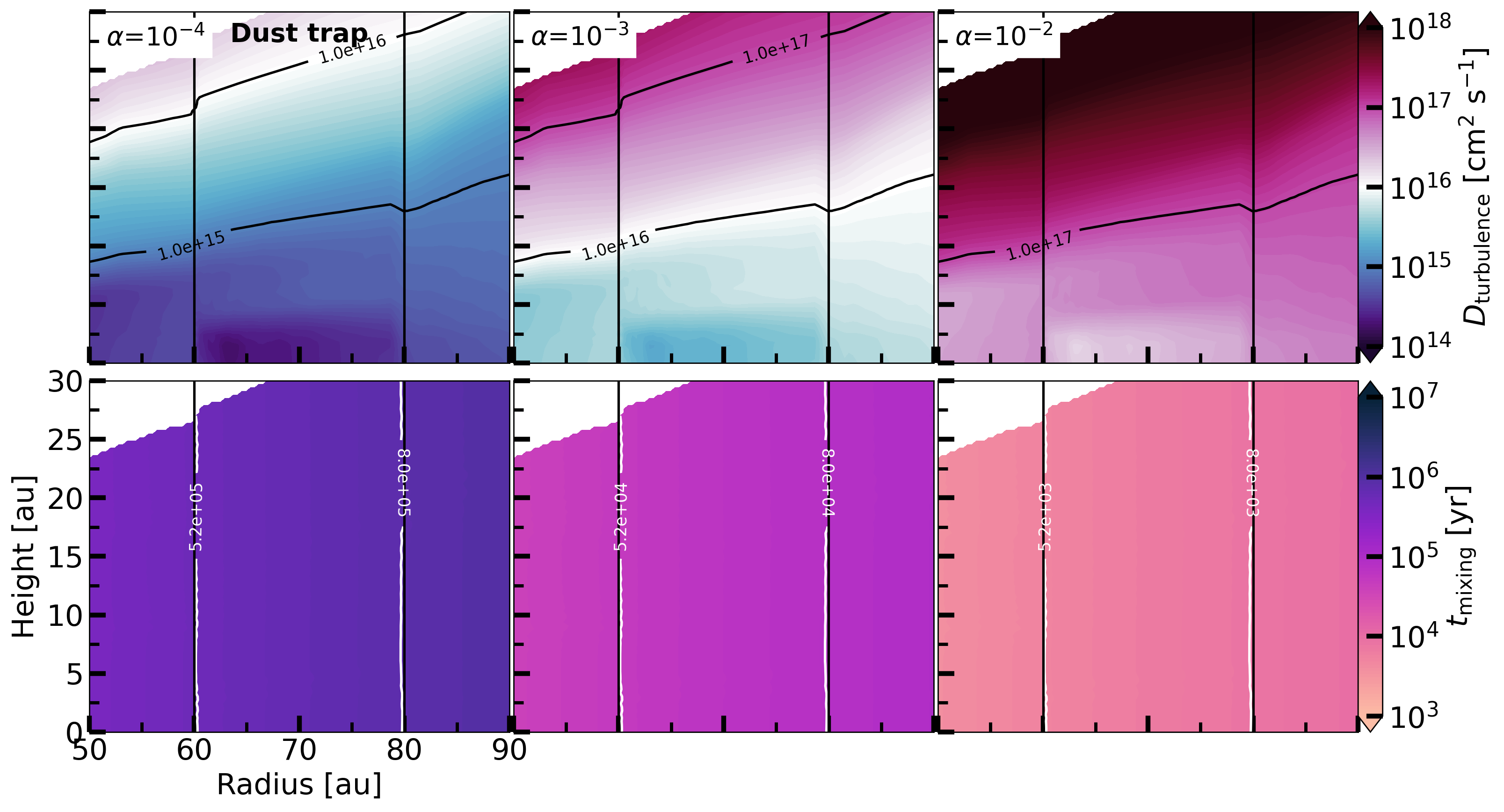}
    \caption{The turbulent diffusion coefficient (in cm$^{2}$~s$^{-1}$, top panel) and the vertical mixing timescale (bottom panel) according to the DALI model of \citet{LeemkerEA23}. The coefficient and timescale have been calculated for $\alpha$ values of 10$^{-4}$ (left panels), 10$^{-3}$ (middle panels), and 10$^{-2}$ (right panels).}
    \label{fig:Model-Diffusion}
\end{figure*}

\begin{figure*}[ht!]
    \centering
    \includegraphics[width=\textwidth]{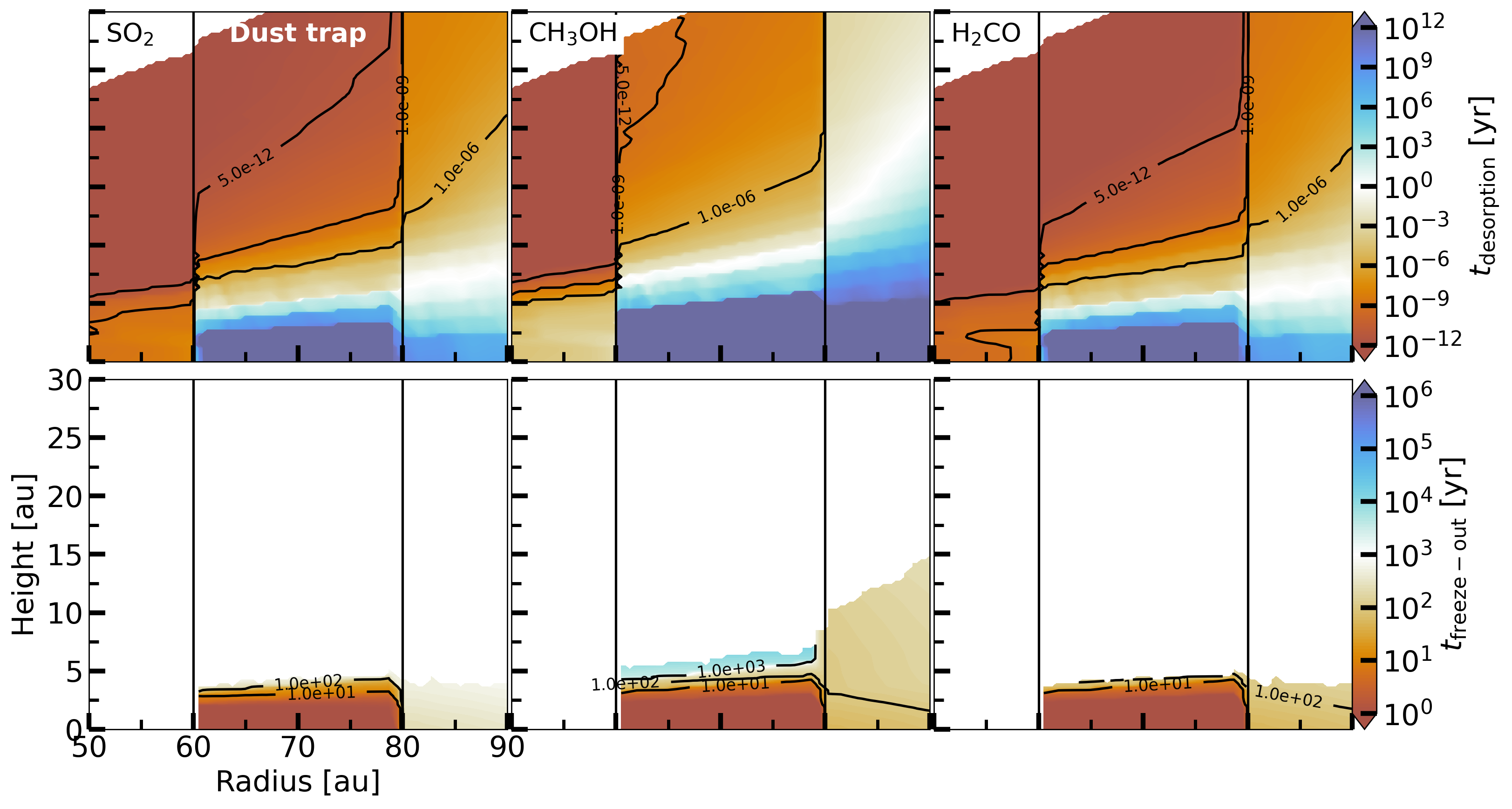}
    \caption{The desorption (top panels) and freeze-out (bottom panels, only showing $t_\textnormal{freeze-out}<t_\textnormal{desorption}$) timescales for \ce{SO_2} (left panels), \ce{CH_3OH} (middle panels), and \ce{H_2CO} (right panels) following the DALI model of \citet{LeemkerEA23}.}
    \label{fig:Model-FD}
\end{figure*}

\begin{figure*}[ht!]
    \centering
    \includegraphics[width=\textwidth]{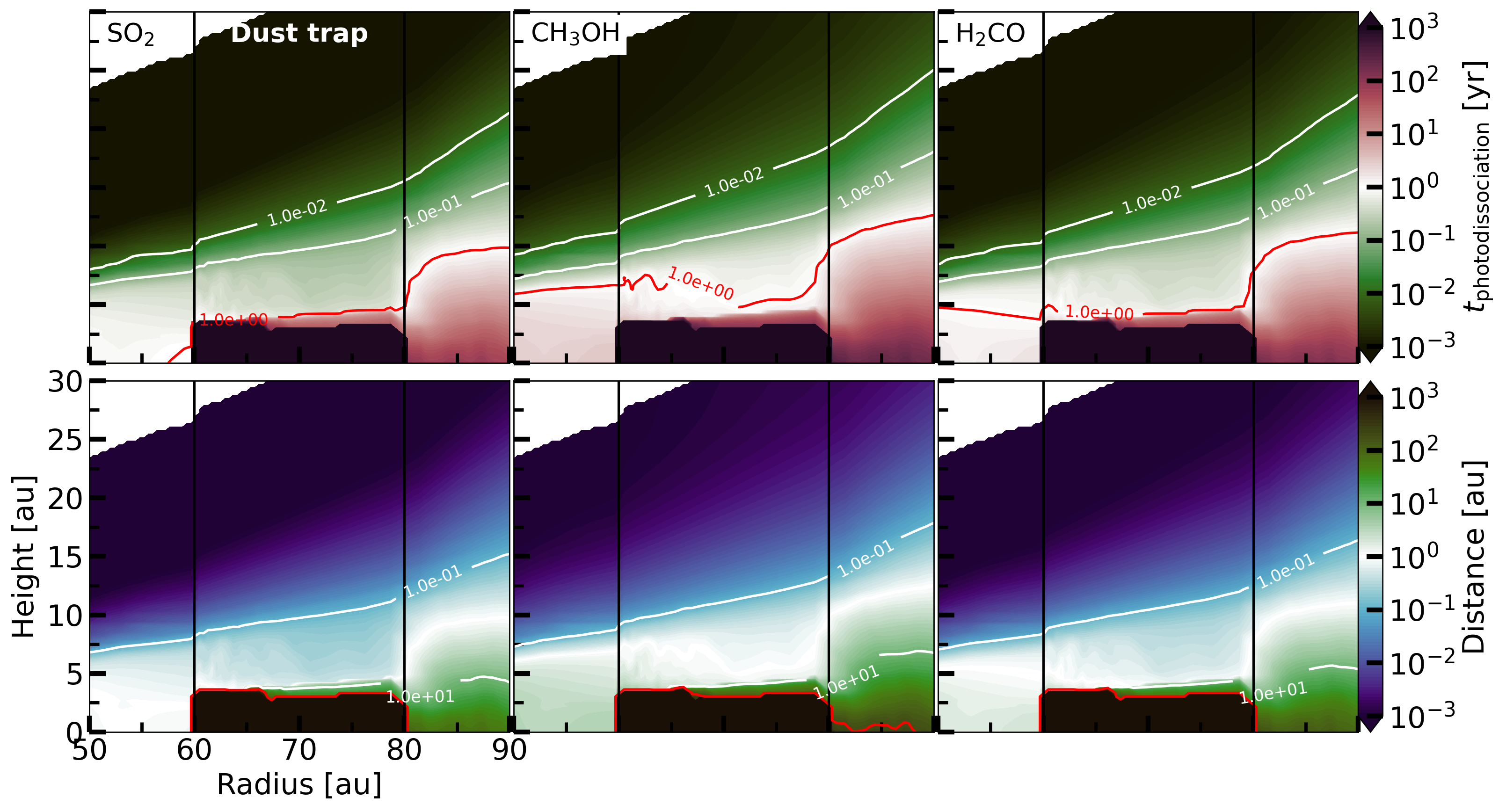}
    \caption{The photodissociation timescales (top panels) and the azimuthal distances (by Keplerian motion) the molecules can travel before they get dissociated (bottom panels) based on the DALI model of \citet{LeemkerEA23}. From left to right, the timescales and distances are shown for \ce{SO_2}, \ce{CH_3OH}, and \ce{H_2CO}, respectively.}
    \label{fig:Model-PD}
\end{figure*}

\end{appendix}

\end{document}